\documentclass[floatfix,superscriptaddress,a4paper,
               nofootinbib,preprint]{revtex4}

\pdfoutput=1

\usepackage{graphicx}
\usepackage{amsmath}
\usepackage{amssymb}

\newcommand{\hsp}{\hspace*{1pt}}
\newcommand{\hspm}{\hspace*{.5pt}}
\newcommand{\ds}{\displaystyle}
\newcommand{\be}{\begin{equation}}
\newcommand{\ee}{\end{equation}}
\newcommand{\bel}[1]{\be\label{#1}}
\newcommand{\re}[1]{Eq.~(\ref{#1})}

\begin{document}

\title{Phase transitions and Bose-Einstein condensation in~alpha-nucleon matter
}

\author{L.~M. Satarov}
\affiliation{
Frankfurt Institute for Advanced Studies, D-60438 Frankfurt am Main, Germany}
\affiliation{
National Research Center ''Kurchatov Institute'' 123182 Moscow, Russia}

\author{I.~N. Mishustin}
\affiliation{
Frankfurt Institute for Advanced Studies, D-60438 Frankfurt am Main, Germany}
\affiliation{
National Research Center ''Kurchatov Institute'' 123182 Moscow, Russia}

\author{A.~Motornenko}
\affiliation{
Frankfurt Institute for Advanced Studies, D-60438 Frankfurt am Main, Germany}
\affiliation{
Institut f\"ur Theoretische Physik,
Goethe Universit\"at Frankfurt, D-60438 Frankfurt am Main, Germany}

\author{V.~Vovchenko}
\affiliation{
Frankfurt Institute for Advanced Studies, D-60438 Frankfurt am Main, Germany}
\affiliation{
Institut f\"ur Theoretische Physik,
Goethe Universit\"at Frankfurt, D-60438 Frankfurt am Main, Germany}

\author{M.~I. Gorenstein}
\affiliation{
Frankfurt Institute for Advanced Studies, D-60438 Frankfurt am Main, Germany}
\affiliation{
Bogolyubov Institute for Theoretical Physics, 03680 Kiev, Ukraine}

\author{H. Stoecker}
\affiliation{
Frankfurt Institute for Advanced Studies,
D-60438 Frankfurt am Main, Germany}
\affiliation{
Institut f\"ur Theoretische Physik,
Goethe Universit\"at Frankfurt, D-60438 Frankfurt am Main, Germany}
\affiliation{
GSI Helmholtzzentrum f\"ur Schwerionenforschung GmbH, D-64291 Darmstadt, Germany}

\begin{abstract}
The equation of state and phase diagram of isospin-symmetric chemically equilibrated mixture
of alpha particles $\alpha$ and nucleons $N$ are studied in the mean-field approximation.
The model takes into account the effects of Fermi and Bose statistics for $N$
and $\alpha$, respectively.
We use Skyrme-like parametrization of the mean-field potentials as functions of
partial densities $n_\alpha$ and~$n_N$, which contain both attractive and repulsive terms. Para\-meters of these potentials are chosen by fitting known
properties of pure $N$- and pure \mbox{$\alpha$-matter} at zero temperature.
The sensitivity of results to the choice of the $\alpha N$ attraction strength is investigated.
The~phase diagram of
the $\alpha-N$ mixture is studied with a special attention paid to
the liquid-gas phase transitions and the Bose-Einstein condensation of $\alpha$ particles.
We~have found two first-order phase transitions, stable and metastable, which
differ significantly by the~fractions of alpha particles. It is shown that states with
alpha condensate are metastable.
\end{abstract}

\maketitle

\section{Introduction}

At subsaturation densities and low temperatures nuclear matter has a tendency
for clusterization, when small and big nucleon clusters are formed under
conditions of thermal and chemical equilibrium. This state of excited nuclear
matter is realized in nuclear reactions at intermediate energies known as
multi-fragmentation of nuclei~\cite{Bon76}. It is believed that clusterized nuclear matter is also
formed in outer regions of neutron-stars and in supernova explosions~\cite{Lat91}.
It may play an important role by providing ''seed'' nuclei for later nucleosynthesis.

Different models have been used to describe the clusterized nuclear matter.
In particular, the statistical approach turned out to be very successful to explain
the mass and energy distributions of fragments and hadrons produced in heavy--ion collisions,
see e.g. Refs.~\cite{Sto83,Bon85}. Another powerful method is to
perform molecular-dynamical simulations in a box taking into account effective
interactions between nucleons, as it was done, e.g., in Ref.~\cite{Pei91}.

To better understand properties of clusterized nuclear matter one should use more
rea\-listic interactions between different clusters and take into account phenomenological constraints.
In our recent paper~\cite{Sat17} we studied the equation of state (EoS) of an idealized
system composed entirely of $\alpha$-particles. Their interaction was described by
a~Skyrme-like mean-field potential. We have found that such a system exhibits two interesting phenomena, namely, the Bose-Einstein condensation (BEC) and the liquid-gas phase transition (LGPT). \mbox{Earlier} the cold alpha matter has been considered microscopically by using phenomeno\-logical $\alpha\alpha$~potentials in~Refs.~\cite{Cla66,Joh80,Bri73}, the lattice calculations were made in~Ref.~\cite{Sed06} and the~relati\-vistic mean-field~(RMF) approach was applied in Ref.~\cite{Mis17}. Properties of cold~$\alpha$~chains have been discussed in Refs.~\cite{Hor72,Mis84,Toh18}.

However, by introducing such one-component system one disregards a possible dissociation
of alphas into lighter clusters and nucleons. This process should be important at nonzero temperatures and large enough baryon densities.
Binary $\alpha-N$ matter in chemical equilibrium with respect to reactions $\alpha\leftrightarrow 4N$ has been considered in~\cite{Hor06} by using the~virial
approach. One should have in mind that the results of~Ref.~\cite{Hor06} may be justified only at small baryon densities. The two-component van der Waals model with excluded-volume repulsion has been developed to describe properties of $\alpha-N$ mixture in Ref.~\cite{Vov17a}. Note that both these approaches disregard possible BEC phenomena.

The EOS of matter composed of nucleons and nuclear clusters have been considered within different approaches including the liquid-drop model~\cite{Lat91}, several versions of the statistical model~\cite{Bot10,Buy14,Fur18} and the RMF models~\cite{Hem10,Pai18,Typ18}. In particular, in~Ref.~\cite{Pai18} the RMF calculations have been performed  with
the~medium-dependent binding energy of alphas.
Comparison of the~excluded-volume and virial EoSs has been made in Ref.~\cite{Lal18}.
However, all these models do not include a possibility of~BEC. This phenomenon was considered
within the quasiparticle approach of~Ref.~\cite{Sed17}, but only for dilute (nearly ideal-gas) mixtures of nucleons and nuclear clusters.

In the present paper we consider the isospin-symmetric $\alpha-N$ matter under the conditions
of chemical equilibrium. The EoS of such matter is calculated in the mean-field approach using
Skyrme-like parametrizations of the mean-field potentials. In this study we
simultaneously take into account the LGPT and BEC effects.

The article is organized as follows.
In Sec.~\ref{lab0} we formulate main features of the model. The limit of ideal $\alpha-N$ gas is
considered in Sec.~\ref{lab1} and Appendix A. Pure nucleon and pure alpha matter with Skyrme interactions are studied in Sec.~\ref{lab2} and \ref{lab3}, respectively. The~results of these sections are used in choosing parameters of mean fields for $\alpha-N$ matter in~Sec.~\ref{lab4a}. The~EoS and phase transitions of such matter are studied numerically in Sec.~\ref{lab6a}. Finally, the conclusions and outlook are given in Sec.~\ref{lab8}.

\section{General remarks and limiting cases}
\vspace*{-2mm}
\subsection{Chemical equilibrium conditions\label{lab0}}

Let us consider the iso-symmetric system (with equal numbers of protons and neutrons)
composed of nucleons $N$ and alpha-particles $\alpha$.
A small difference between the proton and neutron masses and the Coulomb
interaction effects will be neglected.
Our consideration will be restricted to small temperatures $T\lesssim 30$~MeV.
In this case, production of pions and other mesons, as well as excitation of baryonic resonances,
like $\Delta$ and $N^*$, become negligible. Besides, the masses $m_N\simeq 938.9$~MeV and $m_\alpha\simeq 3727.3$~MeV
are much larger than the~system temperature, thus, a non-relativistic approximation
can be used in the~lowest order in $T/m_N$.

In the grand canonical ensemble the pressure $p\,(T,\mu)$
is a function of temperature~$T$ and baryon chemical potential  $\mu$.
The latter is responsible for conservation of the baryon charge.
The chemical potential
of $N$ and $\alpha$ satisfy the relations
\bel{cce}
\mu_N~=~\mu~,~~~~~~\mu_\alpha~=~4\mu~,
\ee
which correspond to the condition of chemical equilibrium in the $N-\alpha$ mixture
due to reactions~\mbox{$\alpha\leftrightarrow 4N$}.
The baryonic number density~$n_B\hsp (T,\mu)=n_N+4\hspm n_\alpha$, the entropy density~$s\hsp (T,\mu)$,
and the energy density $\varepsilon\hsp (T,\mu)$ can be calculated from~$p\,(T,\mu)$ as
\bel{therm}
n_B=\left(\frac{\partial p}{\partial \mu}\right)_T,~~~~ s=\left(\frac{\partial p}{\partial T}\right)_{\ds\mu},
~~~~\varepsilon=T s+\mu\hsp n_B-p
\ee
in the thermodynamic limit, where the system volume goes to infinity.

\vspace*{3mm}
\subsection{Ideal gas limit\label{lab1}}

Let us first consider the $\alpha-N$ system as a mixture of the ideal Fermi-gas of nucleons
and the ideal Bose-gas of alpha.
The pressure of such a system is equal to the sum of partial pressures
\bel{prid}
p^{\,\rm id}(T,\mu)~=~p^{\,\rm id}_N(T,\mu_N)~+p^{\,\rm id}_{\alpha}(T,\mu_\alpha)\,.
\ee
Here ($\hbar=c=k_B=1$):
\bel{prfb}
p^{\,\rm id}_i(T,\mu_i)~=~\frac{g_i}{(2\pi)^3}\int d^{\,3}k\frac{k^2}{3E_i}\left[
\exp{\left(\frac{E_i-\mu_i}{T}\right)}\pm 1\right]^{-1}~~~~(i=N,\alpha)\hsp ,
\ee
where $E_i=\sqrt{m_i^2+k^2}$ and $g_i$ is the spin-isospin degeneracy factor ($g_\alpha=1, g_N=4$).
Upper and lower signs in~\re{prfb} correspond to~$i=N$ and~$i=\alpha$, respectively.

By taking derivatives with respect to $\mu_i$ one gets the partial densities
\bel{denfb}
n_i=\left(\frac{\partial\hsp p^{\,\rm id}_i}{\partial\hsp\mu_i}\right)_T=
\frac{g_i}{(2\pi)^3}\int d^{\,3}k\left[
\exp{\left(\frac{E_i-\mu_i}{T}\right)}\pm 1\right]^{-1}~~~~~(i=N,\alpha)\hsp .
\ee
In the following we will also  use
the canonical variables $T,n_i$ as independent quantities. The~transition from the grand
canonical variables $T,\mu_i$ is made by solving the transcendental equations~(\ref{denfb})
with respect to $\mu_i$. Allowable states of chemically equilibrated
$\alpha-N$ mixture is then found using \re{cce}.

The chemical potential of $\alpha$ particles is restricted by the relation
$\mu_\alpha\leqslant m_\alpha$.
At~$\mu_\alpha=m_\alpha$ the Bose condensation of $\alpha$'s occurs. In this case
a nonzero density of Bose-condensed (zero-momentum)
alpha particles,~$n_{\rm bc}$, should be taken into account. By taking the lowest order approximation
in $T/m_i$ (see Appendix) and introducing the thermal wave length  of~$i$\hsp th particle,
\mbox{$\lambda_i(T)=(2\pi/m_iT)^{1/2}$}, one gets the following relations for
the total density and pressure of~$\alpha$'s in the region of~BEC:
\bel{dabec}
n_\alpha=n_\alpha^*(T)+n_{bc},~~~p_\alpha^{\hsp\rm id}=p_\alpha^*(T)~~~~~(\mu_\alpha=m_\alpha)\hsp .
\ee
Here
\bel{npbec}
n_\alpha^*(T)=n_\alpha(T,{\mbox{$\mu_{\alpha}\to m_\alpha-0$}})\simeq
\frac{g_\alpha}{\lambda^3_\alpha(T)}\hsp\zeta(3/2)\,,
~~~p_\alpha^*(T)\simeq\frac{g_\alpha T}{\lambda^3_\alpha(T)}\hsp\zeta(5/2)\,,
\ee
where $\zeta(x)=\sum_{k=1}^{\infty}k^{-x}$ is the Riemann zeta function
($\zeta(3/2)\simeq 2.612,~~\zeta(5/2)\simeq 1.341$)\,.

\begin{figure}[ht!]
\centering
\includegraphics[trim=1.7cm 7.2cm 2cm 8.5cm, clip, width=0.75\textwidth]{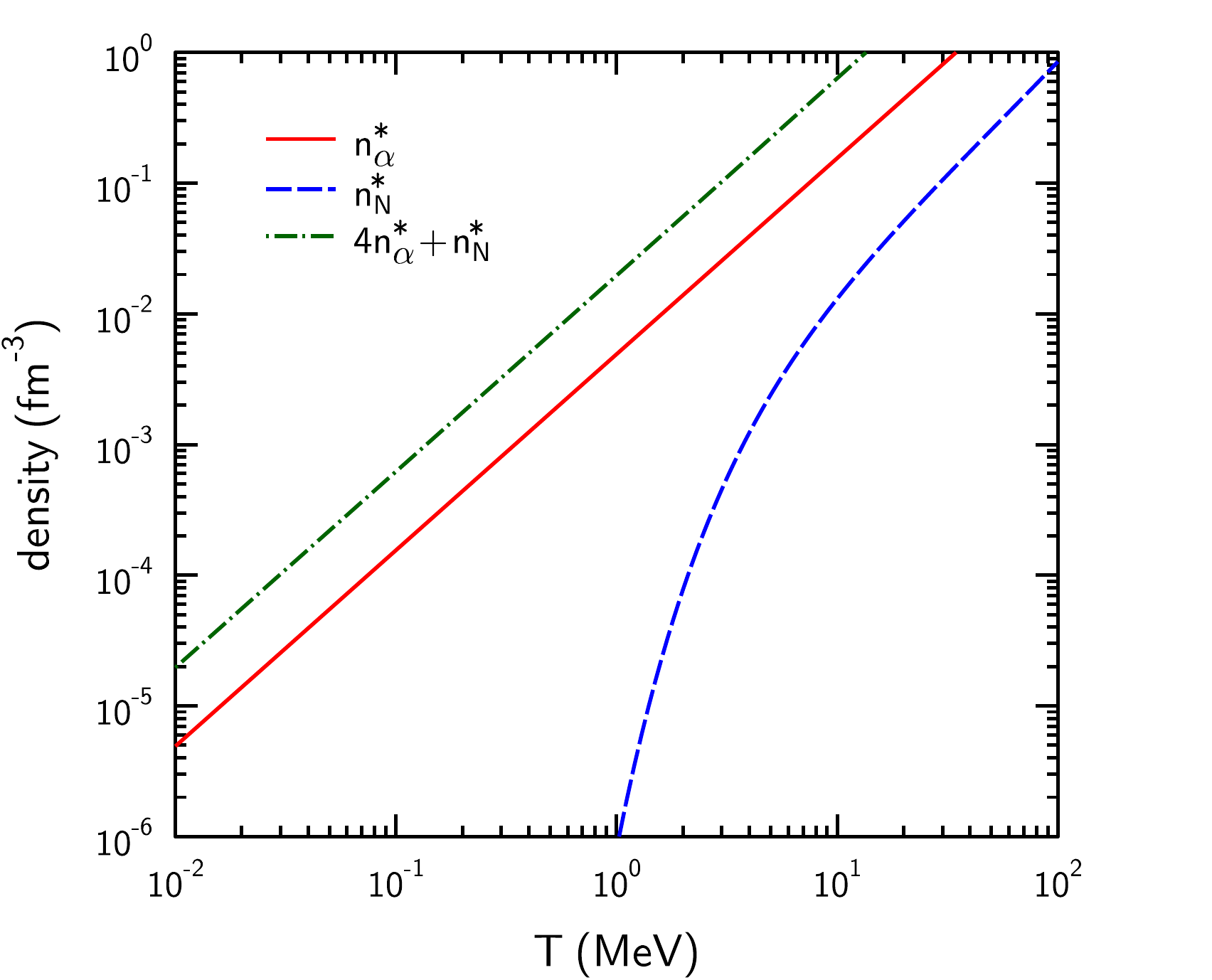}
\caption{
Densities of alphas (the solid line), nucleons (the dashed curve), and
the baryon density~(the~dash-dotted line) at BEC boundary as functions
of temperature for ideal $\alpha-N$ gas.
}\label{fig1}
\end{figure}

The condition (\ref{cce}) leads to the relations
\mbox{$\mu\equiv \mu_N=\mu_\alpha/4\leqslant m_\alpha/4=m_N-B_\alpha$},
where \mbox{$B_\alpha=m_N-m_\alpha/4\simeq 7.1~\textrm{MeV}$} is the binding energy per
baryon of the $\alpha$ nucleus. At~\mbox{$\mu=m_N-B_\alpha$} the BEC occurs in the ideal gas. Using~\re{A1} one obtains the nucleon density in the BEC region
\bel{dnb1}
n_N=n_N^*(T)\simeq\frac{g_N}{\lambda^3_N(T)}\hsp\Phi^+_{3/2}\left(-\frac{B_\alpha}{T}\right),
\ee
where $\Phi^+_{3/2}(\eta)$ is a dimensionless function defined in Appendix. Note that
$n_N^*$ does not depend on $n_{\rm bc}$. From~Eqs.~(\ref{npbec}) and (\ref{dnb1}), one can
get the following relations for the ideal~$\alpha-N$ gas in the BEC domain
\bel{denr}
\frac{n_N}{n_\alpha}\leqslant\frac{n_N^*}{n_\alpha^*}<\frac{1}{2\hsp\zeta(3/2)}e^{-B_\alpha/T}.
\ee
Here we take into account that \mbox{$\Phi^+_{3/2}(\eta)<e^{\hsp\eta}$} and
\mbox{$\lambda_\alpha/\lambda_N\simeq 1/2$}.
According to (\ref{denr}), the fraction of unbound nucleons is small in the whole BEC region,
especially at low temperatures (this conclusion has been earlier made in Ref.~\cite{Sed17}).
Figure~\ref{fig1} shows $n^*_\alpha,n^*_N$ as well as the baryon density of the ideal gas
at the BEC boundary
as functions of $T$. One can see that $n^*_N\ll n^*_\alpha$ even at large temperatures.

\begin{figure}[ht!]
\centering
\includegraphics[trim=1.7cm 7.2cm 2cm 8.5cm, clip, width=0.75\textwidth]{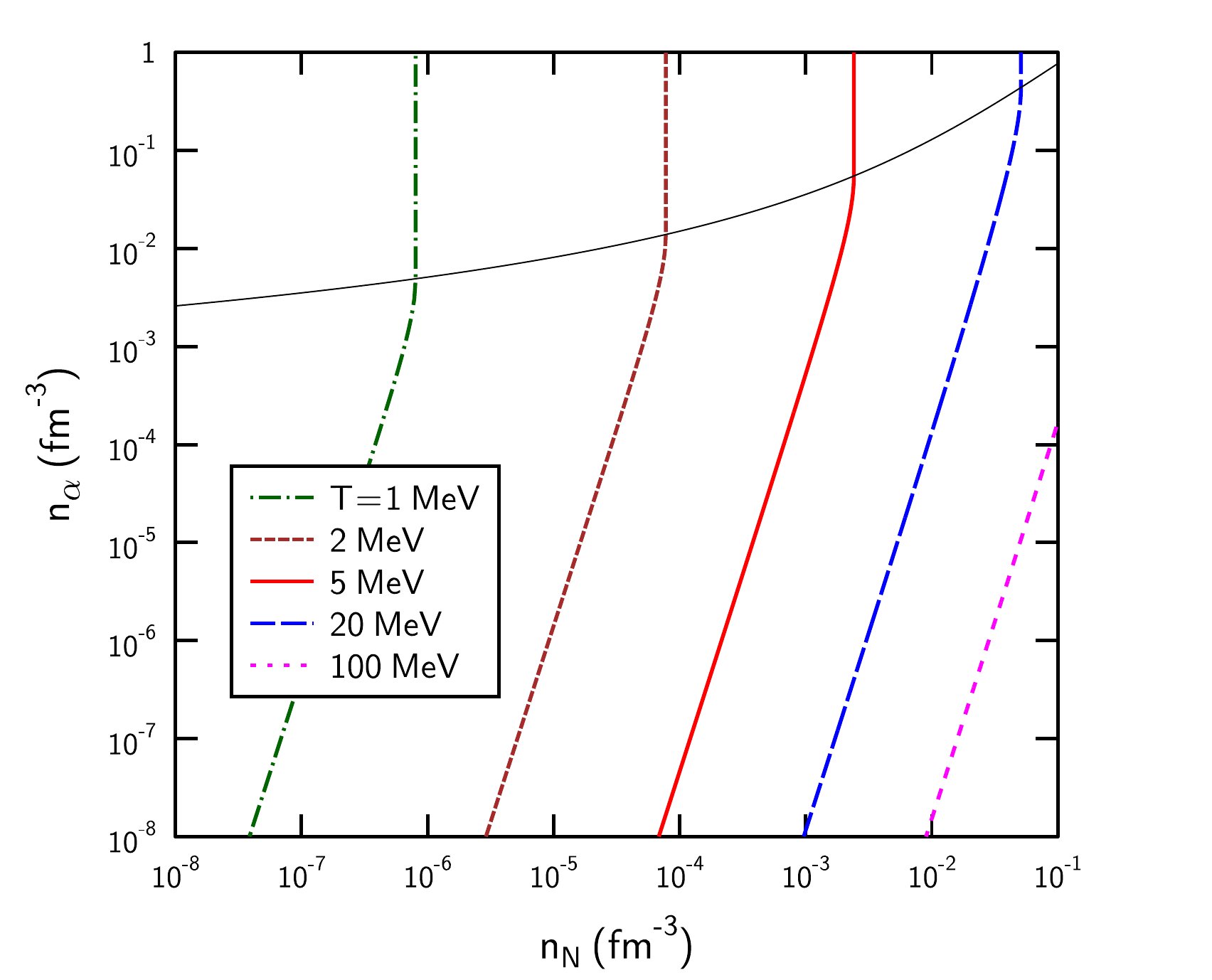}
\caption{
Isotherms of chemical equilibrium for ideal $\alpha-N$ gas.
The thin solid line shows boundary of BEC region.
}\label{fig2}
\end{figure}

At each temperature, the chemical equilibrium condition~(\ref{cce}) gives
a line of allowable states in the ($n_N,n_\alpha$) plane. These lines  are shown in
Fig.~\ref{fig2}. Vertical sections of the
lines correspond to the BEC states with densities $n_N=n^*_N(T)$ and $n_\alpha>n^*_\alpha(T)$.
Such states lie above the BEC boundary
shown by the thin solid curve in Fig.~\ref{fig2}.

\subsection{Pure nucleon matter\label{lab2}}

Let us consider the limiting case of the one-component, iso-symmetric nucleon matter with interaction.
The EoS and the phase diagram of a pure $N$-matter were~studied by many authors.
In particular, the mean-field approximation has been applied in~Refs.~\cite{Anc92,Anc94,Sat09}.
In~such an approach, one introduces a shift of the chemical potential~$\mu_N$
with respect~to the~ideal nucleon gas. We apply the equation
\bel{cpmf}
\widetilde{\mu}_N=\mu_N-U_N(n_N)\hsp,
\ee
where $U_N(n_N)$ is the mean-field potential of nucleons and
$\widetilde{\mu}_N=\widetilde{\mu}_N(T,n_N)$ is the effective
chemical potential of nucleons at the density $n_N$ and temperature $T$. This quantity
is determined by solving~\re{denfb} with $i=N$ and $\mu_i=\widetilde{\mu}_N$\hsp.
Here and below we neglect a~possible explicit dependence of the mean-field potential on temperature. Equation~(\ref{cpmf}) leads to the expression
\bel{psmf}
\Delta p_N~\equiv~ p_N(T,\mu_N)~-~p_N^{\hsp\rm id}(T,\widetilde{\mu}_N)~=~n_N\hsp U_N\hsp(n_N)~-~
\int\limits_0^{n_N}dn_1\hsp U_N(n_1)
\ee
for the shift of the nucleon pressure with respect to its ideal gas value\hsp\footnote
{
This shift is often called as the 'excess' pressure~\cite{Anc92,Anc94}\hspm.
}.
One can see that $\Delta p_N$
does not depend on $T$. From Eqs.~(\ref{cpmf}) and (\ref{psmf}) one can prove validity of
the thermodynamic consistency relation,~\mbox{$n_N=(\partial p_N/\partial\hsp\mu_N)_T$}\hsp.

Further on we use the Skyrme-like parametrization~\cite{Sat09} of the mean field
\bel{smmp}
U_N(n_N)~=~-~2a_N\hsp n_N~+~\frac{\gamma+2}{\gamma+1}\,b_N\hsp n_N^{\gamma+1}\hsp,
\ee
where positive constants $a_N,b_N,\gamma$ are adjustable parameters.
The first and second terms describe, respectively, contributions of medium-range
attractive and short-range repulsive interactions of nucleons.
Substituting~(\ref{smmp}) into~\re{psmf} one obtains
\bel{smdp}
\Delta p_N~=~-~a_N\hsp n_N^2~+~b_N\hsp n_N^{\gamma+2}\hsp.
\ee

Parameters entering Eqs.~(\ref{smmp}) and (\ref{smdp}) are chosen to reproduce
phenomenological pro\-perties of equilibrium iso-symmetric nuclear matter at $T=0$\hsp. We use the values~\cite{Sat09}
\bel{eqnm}
\textrm{min}\hsp\left(\frac{E}{B}\right)=-15.9~\textrm{MeV},~~n_N=n_0=0.15~\textrm{fm}^{-3}~~~~~(T=0)
\ee
for the binding energy per baryon, $E/B=\varepsilon_N/n_N-m_N$, and the saturation density~$n_0$\hsp.
Using further the thermodynamic identities at zero temperature,
\mbox{$p_N=n_N^2d\hsp (\varepsilon_N/n_N)/dn_N$} and~\mbox{$\mu_N=(\varepsilon_N+p_N)/n_N$},
one finds the equations which are equivalent to (\ref{eqnm}):
\bel{eqnm1}
\mu_N=\mu_0=923~\textrm{MeV}\hsp,~~p_N=0~~~~~(T=0,~n_N=n_0)\hsp .
\ee

At $T\to 0$ one can calculate the integrals in Eqs.~(\ref{prfb}) and (\ref{denfb})
for~$i=N$ analytically. In~this limit the Fermi distributions inside these
integrals can be replaced by unity if $k<k_F$ where $k_F=(6\hsp\pi^2n_N/g_N)^{1/3}$
is the Fermi momentum of nucleons. One gets the relations
\bel{mupzt}
\widetilde{\mu}_N(T=0,n_N)=E_F(n_N)=\sqrt{k_F^2+m_N^2}~,
~~~~p_N^{\hsp\rm id}(T=0,n_N)=\frac{g_N}{6\pi^2}\int\limits_0^{k_F}\frac{k^4\hsp dk}{\sqrt{k^2+m_N^2}}\,.
\ee
From Eqs.~(\ref{cpmf})--(\ref{psmf}) and (\ref{eqnm1})--(\ref{mupzt}) one obtains two
equations
\bel{eqnm2}
U_N(n_0)=\mu_0-E_F(n_0)~,~~~~\Delta p_N(n_0)=-\hsp p_N^{\hsp\rm id}(T=0,n_0)
\ee
for the parameters $a_N,b_N$ as functions of $\gamma$\hsp .

The results of numerical calculation for $\gamma=1/6$ and $\gamma=1$
are shown in Table~I. In~addi\-tion to coefficients of the Skyrme interactions,
we also present the values of the~incompressi\-bi\-lity modulus
\bel{imcm}
K_N=9\frac{d\hsp p_N}{d\hsp n_N}=9\hsp n_N\frac{d\hsp(E_F+U_N)}{dn_N}
\ee
at the saturation point $n_N=n_0,~T=0$\hsp. As noted in Ref.~\cite{Ben03},
the Skyrme-like \mbox{models} with~\mbox{$1/6\leqslant\gamma\leqslant 1/3$} predict
reasonable values of the~\mbox{nuclear} matter compressibi\-lity
\mbox{$K_N=200-240~\textrm{MeV}$}\hsp\footnote
{
See, however,~Ref.~\cite{Sto14} where higher values, $K_N=250-315~\textrm{MeV}$,
have been obtained from the fit of data on the giant monopole resonance.
}.
This agrees with our
calculations. Indeed, one can see from \mbox{Table}~I that the 'soft' Skyrme parametrization
with $\gamma=1/6$ is preferable as compared to~$\gamma=1$.
\begin{table}[ht!]
\caption
{\label{tab1} Characteristics of pure nucleon matter in
Skyrme mean--field model.}
\vspace*{3mm}
\begin{tabular}{|c|c|c|c|c|c|c|}
\hline
~$\gamma$~&~$a_N ~\left(\textrm{GeV\hsp fm}^3\right)$~&~
$b_N ~\left(\textrm{GeV\hsp fm}^{\hsp 3+3\gamma}\right)$~&~$K_N~(\textrm{MeV})$~&~$T_c~(\textrm{MeV})$~
&~\mbox{$n_c~\left(\textrm{fm}^{-3}\right)$}
&~\mbox{$\mu_c-m_N~(\textrm{MeV})$}\\
\hline
~1/6~~&1.167&1.475&198&15.3&0.048&$-31.6$\\
\hline
1&0.399&2.049&372&21.3&0.059&$-42.8$\\
\hline
\end{tabular}
\end{table}

\begin{figure}[ht!]
\centering
\includegraphics[trim=1.7cm 6.6cm 1.7cm 9cm, clip, width=0.48\textwidth]{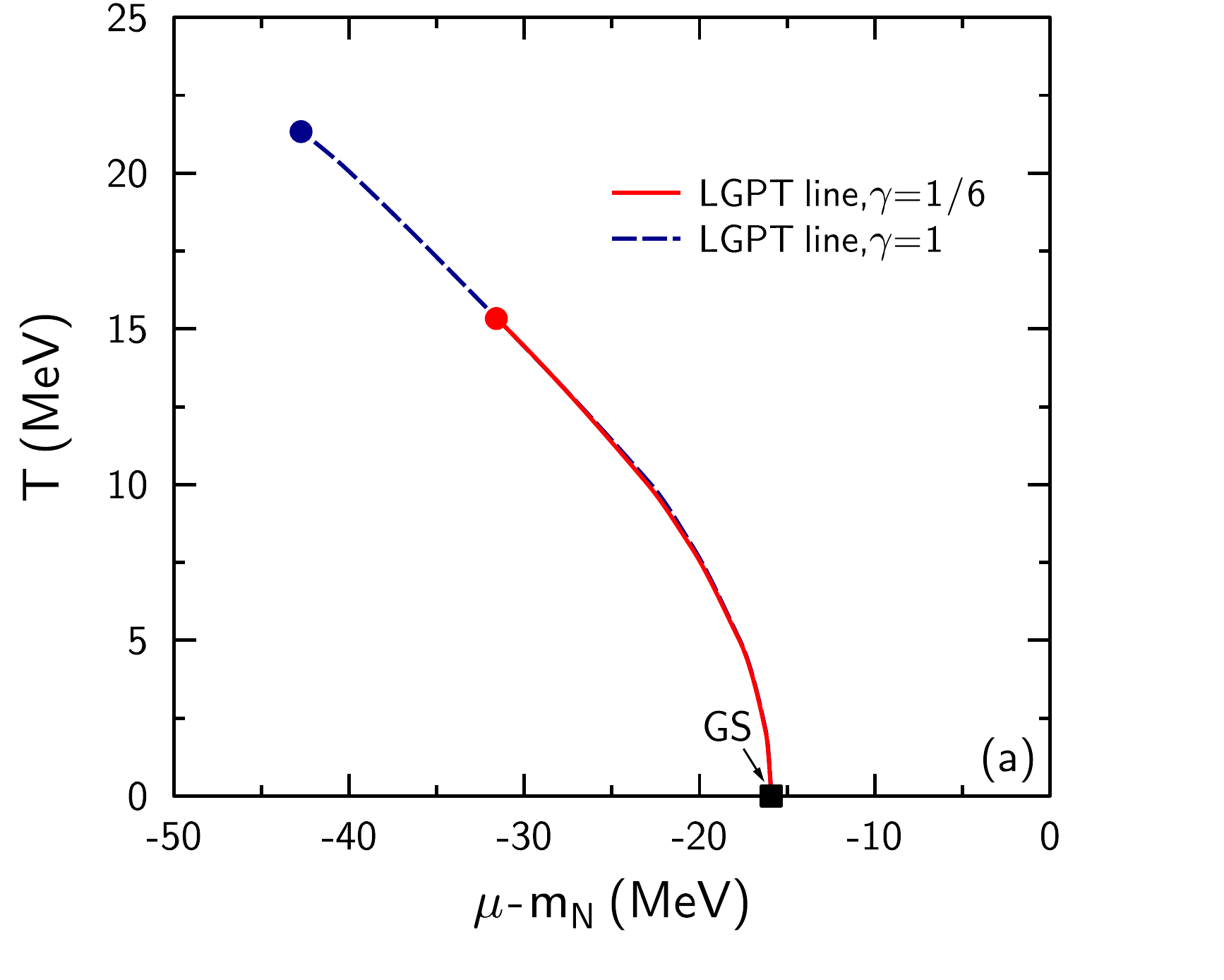}
\includegraphics[trim=1.7cm 6.6cm 1.7cm 9cm, clip, width=0.48\textwidth]{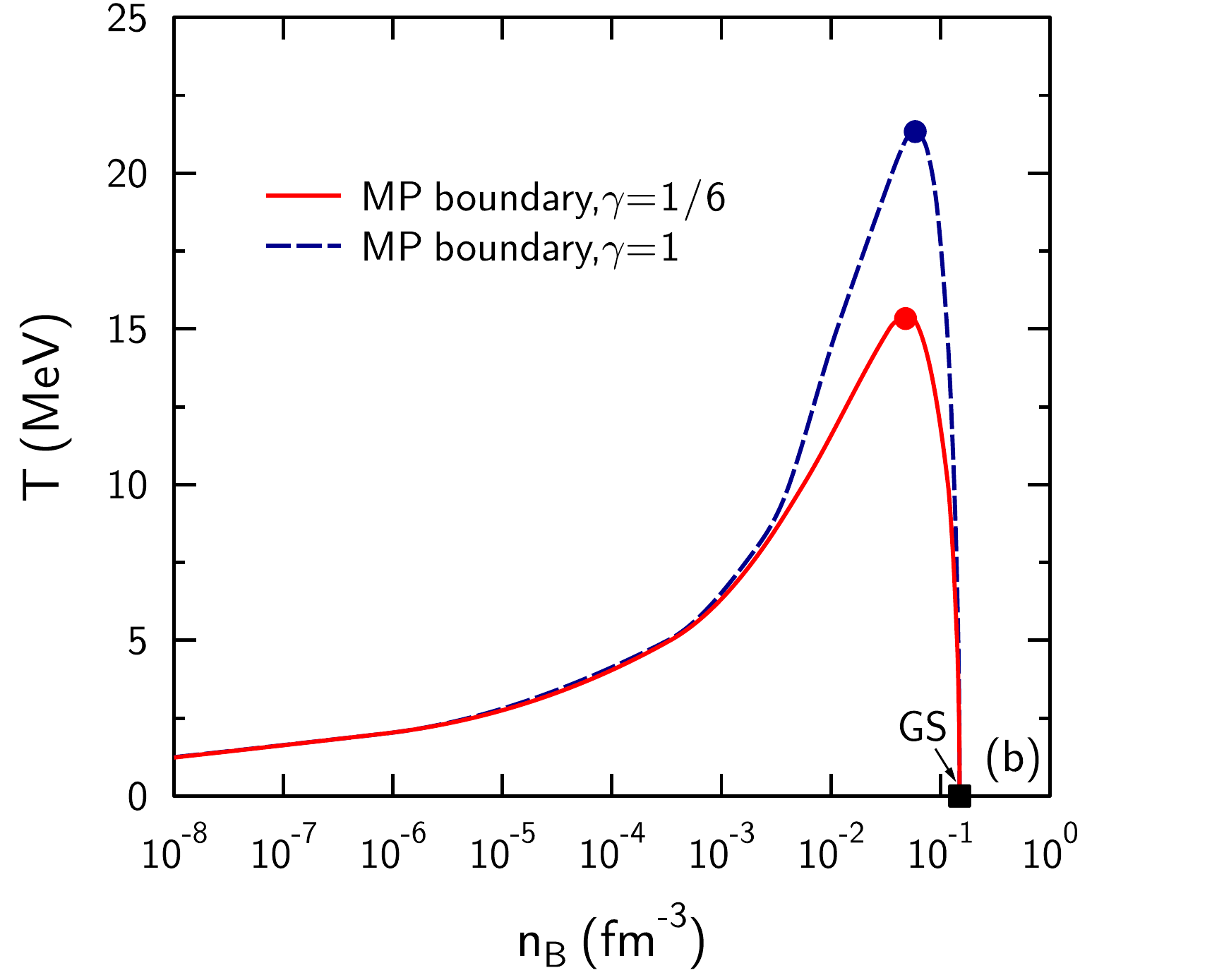}
\caption{
Phase diagrams of iso-symmetric nucleon matter in $(\mu,T)$ (a) and $(n_B,T)$ (b) planes.
The~solid and dashed lines correspond to LGPT at $\gamma=1/6$
and $\gamma=1$, respectively. Full dots mark positions of critical points.
The ground state  is shown by the full square.
}\label{fig3}
\end{figure}
Using Eqs.~(\ref{therm}), (\ref{prfb})--(\ref{denfb}), and (\ref{cpmf})--(\ref{smdp}) one can calculate
all thermodynamic functions of~a~pure nucleon matter at nonzero temperatures.
Our mean-field model predicts a~first-order LGPT
at temperatures $0\leqslant T\leqslant T_c$, where~$T_c$ is the critical temperature. Characte\-ristics of~the~LGPT are found by using the Gibbs conditions of~the
phase equilibrium~\cite{Lan75}. For
isotherms with $T<T_c$ there are two (meta)stable branches of the chemical potential as the function of pressure. In accordance with the Gibbs rule, these branches intersect at the LGPT point. We find the intersection points numerically by calculating isotherms in the chemical potential--pressure plane. Position of the critical point is found by solving two equations~\cite{Lan75}:
\mbox{$(\partial\hsp p_N/\partial\hsp n_N)_T=0$},~\mbox{$(\partial^{\hsp 2} p_N/\partial^{\hsp 2} n_N)_T=0$}. Characteristics of this point for the~soft ($\gamma=1/6$) and stiff ($\gamma=1$) repulsive interaction are given in the last three columns of~Table~I.

Figures~\ref{fig3}\hsp (a) and~(b) show the phase diagrams of nucleonic matter in the $(\mu,T)$ and~$(n_B,T)$ planes,~respectively. The LGPT line in Fig.~\ref{fig3}\hsp (a) goes from the ground state~(GS) at~\mbox{$T=0$},~\mbox{$\mu=\mu_0$} to the
critical point at~\mbox{$T=T_c$},~\mbox{$\mu=\mu_c$}. According to our calcu\-la\-tions,
$T_c$~in\-creases, but $\mu_c$ decreases with $\gamma$. Figure~\ref{fig3}\hsp (b) shows
the 'binodals', i.e., the boundaries of the liquid-gas mixed phase (MP). They intersect
the density axis at $n_B=n_N=n_0$\hspm .

\subsection{Pure alpha matter}\label{lab3}

In this section we consider the idealized case of a pure alpha matter. In~this
limit reactions~\mbox{$\alpha\leftrightarrow 4N$} are disregarded
and, therefore, the chemical equilibrium is~\mbox{violated}.
Up~to now the EoS of such matter is poorly known. The variational microscopic
calculations based on phenomenological $\alpha\alpha$ potentials were
made a long time ago in~Ref.~\cite{Cla66}. More recently the~EoS of a~pure $\alpha$ matter
was considered within several simplified models in Refs.~\cite{Sed06,Mis17}.
In~Ref.~\cite{Sat17} the phase diagram of such matter has been studied within
a~Skyrme mean-field model. Below we apply the same approach and
use characteristics of the $\alpha$-matter~GS obtained in~Ref.~\cite{Cla66}:
\bel{gsalm}
W_\alpha=-\textrm{min}\hsp\left(\frac{E_\alpha}{B}\right)=12~\textrm{MeV}~,
~~~~n_\alpha=n_{0\alpha}=0.036~\textrm{fm}^{-3}~~~~~(T=0)\hsp .
\ee
Note that the baryon density of this state, $4n_{0\alpha}\simeq 0.144~\textrm{fm}^{-3}$, is close to the saturation density of a pure nucleon matter, but the latter has stronger binding per baryon (compare Eqs.~(\ref{eqnm}) and (\ref{gsalm})).

In the case of a pure $\alpha$ matter, one can write the equations, analogous to Eqs.~(\ref{cpmf})--(\ref{smdp})
\bel{mpal}
\widetilde{\mu}_\alpha=4\mu-U_\alpha(n_\alpha)~,~~~~~p_\alpha=p_\alpha^{\hsp\rm id}(T,\widetilde{\mu}_\alpha)+\Delta p_\alpha(n_\alpha)\,,
\ee
where $\mu$ is the baryon chemical potential,
and $U_\alpha$ and $\Delta p_\alpha$ are parameterized by Eqs.~(\ref{smmp}) and (\ref{smdp}) with
the replacement $N\to\alpha$\,. Below we choose the same parameter $\gamma$
as for nucleons and find the coefficients~$a_\alpha$ and $b_\alpha$ from the conditions~(\ref{gsalm}).

In  our mean-field model, one has the following relations for states with the BEC
of alpha particles
\bel{becr1}
\widetilde{\mu}_\alpha=m_\alpha~,~~~~n_\alpha\geqslant n_\alpha^*(T)~,~~~~p_\alpha^{\hsp\rm id}=p_\alpha^*(T)\,,
\ee
where $n_\alpha^*$ and $p_\alpha^*$ are defined in~\re{npbec}. The boundary of the BEC
region is obtained
after replacing the inequality in (\ref{becr1}) by the equality. Solving the resulting equation, \mbox{$4\mu=m_\alpha+U_\alpha\left[\hsp n_\alpha^*(T)\hsp\right]$},
gives a line in the $(\mu,T)$ plane. For brevity, we call it as the BEC line.

At zero temperature $n_\alpha^*=0$ and $p_\alpha^*=0$ and the conditions~(\ref{becr1})
hold for all states. In~this case Eqs.~(\ref{mpal}) give
\bel{zteos}
4\mu=m_\alpha+U_\alpha(n_\alpha)~,~~~~p_\alpha=\Delta\hsp p_\alpha (n_\alpha)~~~~~(T=0)\hsp .
\ee
Using further the relations \mbox{$E_\alpha/B=\varepsilon_\alpha/4n_\alpha-m_N$} and
\mbox{$p_\alpha=4\mu\hsp n_\alpha-\varepsilon_\alpha=0$} for the GS of alpha matter,
one gets algebraic equations for coefficients of the Skyrme interaction~\cite{Sat17}:
\bel{gsal1}
U_\alpha(n_{0\alpha})=4\hsp (B_\alpha-W_\alpha)~,~~~~\Delta\hsp p_\alpha (n_{0\alpha})=0,
\ee
where $B_\alpha$ was introduced in Sec.~\ref{lab1}.
The solutions of Eqs.~(\ref{gsal1}) can be written~as
\bel{saba}
a_\alpha=b_\alpha n_{0\alpha}^\gamma=\frac{4\hsp(\gamma+1)} {~\gamma\hsp n_{0\alpha}}\hsp
(W_{\alpha}-B_\alpha)\,.
\ee

Numerical values of the coefficients $a_\alpha$ and $b_\alpha$ as well as the
compressibility ~\mbox{$K_\alpha=9\hsp\gamma a_\alpha n_{0\alpha}$} are given
in Table~\ref{tab2} for the soft and stiff Skyrme repulsion.
\begin{table}[ht!]
\caption
{\label{tab2} Characteristics of pure alpha matter in
Skyrme model.}
\vspace*{3mm}
\begin{tabular}{|c|c|c|c|c|c|c|c|}
\hline\footnotesize
$\gamma$&\hsp\hsp\mbox{$a_{\hsp\alpha}\left(\textrm{GeV\hsp fm}^3\right)$}
&\hsp\hsp\mbox{$b_{\hsp\alpha}\left(\textrm{GeV\hsp fm}^{\hsp 3+3\gamma}\right)$}&\hsp\hsp \mbox{$K_\alpha\hsp (\textrm{MeV})$}&\hsp\hsp\mbox{$T_c\hsp (\textrm{MeV})$}
&\hsp\hsp\mbox{$n_{Bc}\left(\textrm{fm}^{-3}\right)$}
&\hsp\hsp\mbox{$\mu_c-m_N\hsp (\textrm{MeV})$}
&\hsp\hsp\mbox{$T_{\rm tp}\hsp(\textrm{MeV})$}\\
\hline
~1/6~~&3.831&6.667&207&10.2&0.037&$-16.7$&3.56\\
\hline
1&1.094&30.39&354&13.7&0.048&$-19.3$&3.65\\
\hline
\end{tabular}
\end{table}

\begin{figure}[ht!]
\centering
\includegraphics[trim=1.7cm 6.6cm 1.7cm 8cm, clip, width=0.48\textwidth]{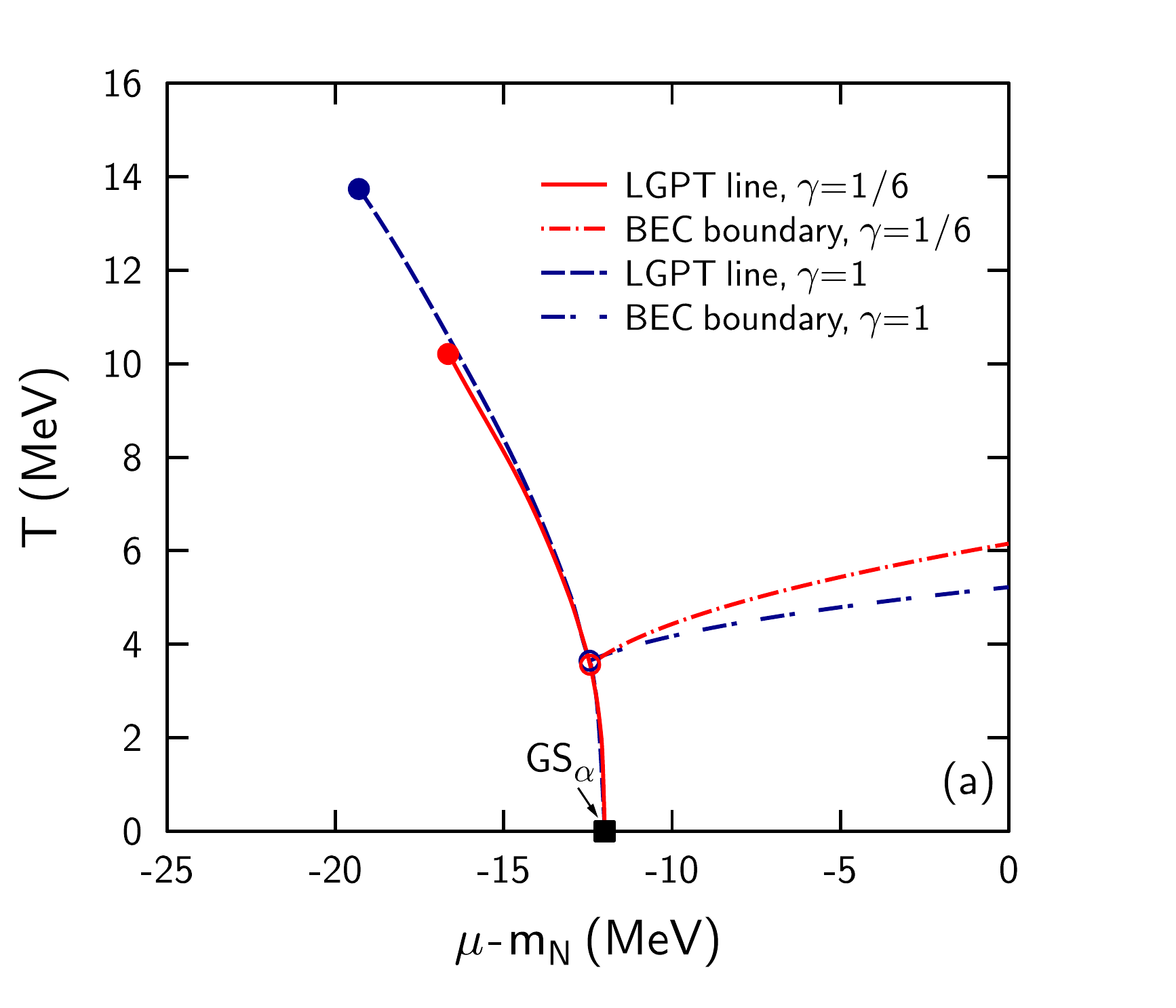}
\includegraphics[trim=1.7cm 6.6cm 1.7cm 8cm, clip, width=0.48\textwidth]{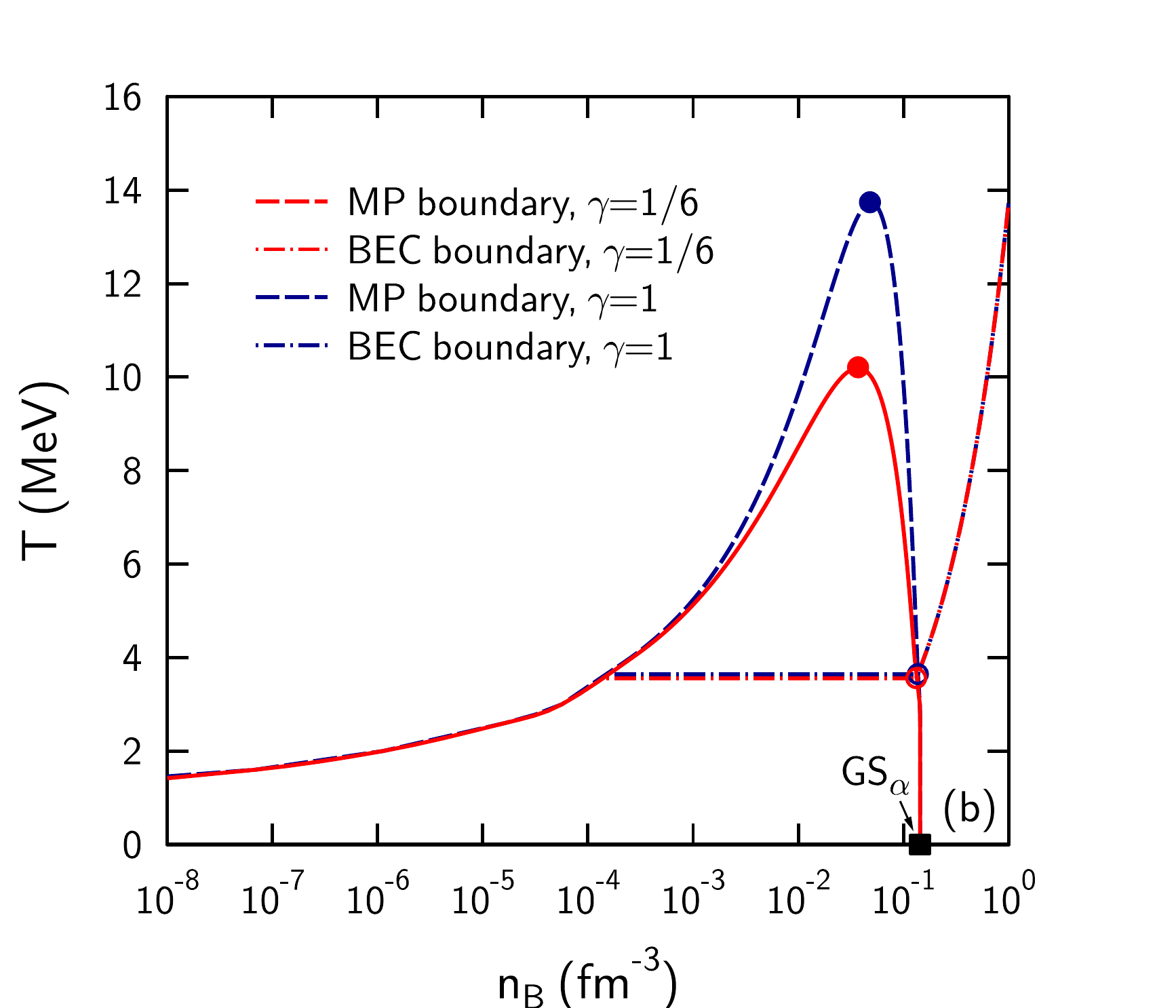}
\caption{
Phase diagrams of pure alpha matter in $(\mu,T)$ (a) and $(n_B,T)$ (b) planes.
The solid and dashed curves correspond to $\gamma=1/6$
and $\gamma=1$, respectively. Full dots mark positions of critical points.
The~GS of alpha matter is shown by the full square.
The dash-dotted lines represent boundaries of~BEC regions.
Open dots show positions of triple points.~They practically coincide for
two considered values of $\gamma$.
}\label{fig4}
\end{figure}
Using the Skyrme interaction
one can calculate the thermodynamical functions of a~pure alpha matter at nonzero temperatures.
Similarly to the pure nucleon matter, the model predicts the LGPT in a pure alpha system.
In the $(\mu,T)$ plane this phase transition occurs along a line  which goes
from the GS at $T=0$ to the critical point at $T=T_c,\mu=\mu_c$.
A~presence of the BEC imposes some complications as compared to the case of
a~pure nucleon matter. We found that
the BEC boundary crosses the LGPT line at some 'triple' point  with
tempera\-ture~$T_{\rm tp}<T_c$.

The resulting phase diagrams in the $(\mu,T)$ and $(n_B,T)$ planes are shown in
Figs.~\ref{fig4}\hsp (a) and~(b), respectively (note that \mbox{$n_B=4\hspm n_{\alpha}$} for
a pure $\alpha$ matter).
Characteristics of the~critical point as well as the temperature of the~triple point
are given in Table~\ref{tab2}.
Similar to~a~pure nucleon matter, the value of $T_c~(\mu_c)$ increases (decreases) with $\gamma$,
but the position of the~triple point
only slightly depends on this parameter. Note that the BEC region in Fig.~\ref{fig4}\hspm (a)
extends to the right from the LGPT line and below the BEC line.
According to Fig.~\ref{fig4}\hspm (b), the~BEC line in the~$(n_B,T)$ plane is not sensitive to $\gamma$ outside the MP region.
It is clear that inside this region the BEC occurs only in the liquid
phase, which volume fraction diminishes with decreasing $n_B$\hspm. Therefore, the
volume fraction of the condensate decreases too and vanishes on the left
binodal boundary. The horizontal lines in Fig.~\ref{fig4}\hspm (b) show the BEC critical
temperatures in the~MP domain for two considered values of~$\gamma$.

\section{Skyrme model for $\alpha-N$ binary mixture\label{lab4a}}

\subsection{Thermodynamic relations for two--component system\label{lab4}}

Similarly to one-component matter, we take into account
multiparticle interactions in the~$\alpha-N$~mixture by introducing a temperature-independent
excess part of pressure~$\Delta p$
\bel{tpwi}
p=p_N^{\hsp\rm id}(T,n_N)+p_\alpha^{\hsp\rm id}(T,n_\alpha)+\Delta p\hsp (n_N,n_\alpha)\hsp .
\ee
A similar expression can be written for the free energy density, $f=\sum\limits_{i=N,\alpha}\mu_i\hsp n_i-p$, after introducing the excess term
\mbox{$\Delta f=f-f_N^{\hsp\rm id}-f_\alpha^{\hsp\rm id}$}.
At known $\Delta p$ one can calculate the mean-field potentials \mbox{$U_i=\mu_i-\widetilde{\mu}_i$} as well as the excess free energy $\Delta f$.
The following relations can be obtained~\cite{Sat15,Sat17a}
\bel{scpf}
\mbox{$U_N\hsp (n_N,n_\alpha)=\left(\frac{\ds\partial\hspm\Delta f}{\ds\partial\hsp n_N}\right)_{n_\alpha}\hspace*{-3pt}$},
~\mbox{$U_\alpha\hsp (n_N,n_\alpha)=\left(\frac{\ds\partial\hspm\Delta f}{\ds\partial\hsp n_\alpha}\right)_{n_N}\hspace*{-3pt}$},
~\mbox{$\Delta f\hsp (n_N,n_\alpha)=\hspace*{-2pt}\int\limits_0^1\frac{\ds d\hsp\lambda}{\ds\lambda^2}
\,\Delta p\hsp (\lambda\hsp n_N,\lambda\hsp n_\alpha)$}\hsp.
\ee
In addition, one can find the entropy density
\mbox{$s=-\partial f/\partial\hsp T=\sum_is_i^{\hsp\rm id}$}~ and the energy density
\mbox{$\varepsilon=f+T\hsp s=\sum_i\varepsilon_i^{\hsp\rm id}+\Delta f$}\,,
where $s_i^{\hsp\rm id}$ and $\varepsilon_i^{\hsp\rm id}$ are the corresponding
ideal-gas quantities for~$i$th species ($i=N,\alpha$).

The free energy density is a~genuine thermodynamic potential in the canonical ensemble.
Instead of partial densities $n_N$ and $n_\alpha$ one can also use the variables
\bel{nvar}
n_B=n_N+4n_\alpha~,~~~~~\chi=\frac{4n_\alpha}{n_B}\hsp .
\ee
The quantity $\chi$ is a fraction of bound nucleons in the $\alpha-N$ matter
(it is approximately equal to the mass fraction of alphas).
Note that due to the baryon number
conservation, \mbox{$B=N_N+4N_\alpha=\textrm{const}$}, the baryon density $n_B$
is inversely proportional to the system \mbox{volume~$V$}. Using~\re{nvar} and thermodynamic
identities, one can write the following relations for changes of the free energy per baryon in any isothermal process
\bel{fenc1}
d\left(\frac{F}{B}\right)=d\left(\frac{f}{n_B}\right)=p\,\frac{d\hsp n_B}{n_B^2}
+\left(\frac{\mu_\alpha}{4}-\mu_N\right)d\hsp\chi~~~~~~(T=\textrm{const})\hsp .
\ee
According to this equation, at fixed $T$ and $n_B$  the quantity $F/B$ reaches its extremum if~the
condition~(\ref{cce}) is satisfied. However, solving~\re{cce}
with respect to $\chi$ does not \mbox{necessarily} gives the true state of chemical
equilibrium. In particular, the solution can be un\-stable~(\mbox{$F/B=\textrm{max}$}) if~the
second derivative of $F/B$ over $\chi$ is negative.

In general, one should explicitly calculate the curvature matrix
$(\partial^{\hsp 2}f/\partial\hsp n_i\hsp\partial\hsp n_j)_T$ to study stability of the system
with respect to fluctuations of partial densities $n_N,n_\alpha$. Only
if both eigenvalues of this matrix are nonnegative, the corresponding state will be
stable\hsp\footnote
{
Note, that appearance of negative-curvature (spinodal) parts of the free--energy density
surface can be considered as a necessary condition for the LGPT.
}.
The~necessary condition of stability can be written as~\cite{Mue95}
\bel{stcon}
\textrm{det}\left(\frac{\partial^{\hsp 2}f}{\partial\hsp n_i\partial\hsp n_j}\right)_T=
\textrm{det}\left(\frac{\partial\hsp\mu_i}{\partial\hsp n_j}\right)_T=
\left(\frac{\partial\hsp\mu_N}{\partial\hsp n_N}\right)_T
\left(\frac{\partial\hsp\mu_\alpha}{\partial\hsp n_\alpha}\right)_T
-\left(\frac{\partial\hsp\mu_N}{\partial\hsp n_\alpha}\right)_T^2\geqslant 0\hsp .
\ee

\subsection{Skyrme parametrization of interaction terms\label{lab5}}

In the present paper we apply a generalized Skyrme--like parametrization
for the excess pressure $\Delta p$:
\bel{dpit}
\Delta p\hsp (n_N,n_\alpha)=-\sum\limits_{i,j}a_{ij}n_i\hsp n_j+\left(\sum\limits_i B_in_i\right)^{\gamma+2},
\ee
where $a_{ij},~B_i$, and $\gamma$ are positive constants and the sums go over $i,j=N,\alpha$.
The first term in the right hand side of~\re{dpit} describes attractive forces and has the same structure as in the two-component van der Waals equation of state~\cite{Vov17a}.
The second term, responsible for repulsive interactions, is obtained by interpolation between the limits $n_\alpha=0$ and $n_N=0$ considered in Sec.~\ref{lab2} and~\ref{lab3}.
From the comparison with these limiting cases one gets
the~relations $a_{\hsp i\hspace*{0.2pt}i}=a_{\hsp i}, B_i^{\hsp\gamma+2}=b_{\hsp i}$ where
$a_{\hsp i}$ and $b_{\hsp i}$ are the Skyrme coefficients introduced earlier for the pure nucleon ($i=N$)
and pure alpha ($i=\alpha$) matter. Using these relations, one finds
\bel{dpit1}
\Delta p\hsp (n_N,n_\alpha)=-(a_N\hsp n_N^2+2a_{N\alpha}n_N\hsp n_\alpha+a_\alpha\hsp n_\alpha^2)
+b_N\hsp (n_N+\xi\hsp n_\alpha)^{\gamma+2},
\ee
where
\bel{cxi}
\xi=\left(\frac{b_\alpha}{b_N}\right)^{1/(\gamma+2)}=
\left\{\begin{array}{ll}
2.006,&\gamma=1/6\hsp ,\\
2.457,&\gamma=1\hsp .
\end{array}\right.
\ee
Numerical values of $\xi$ in~\re{cxi} are obtained by substituting the $b_N$ and
$b_\alpha$ values from Tables~I and II. One can see that there is only one
unknown coefficient in the parametrization~(\ref{dpit1}), namely the cross-term
coefficient $a_{N\alpha}$ which determines the $\alpha-N$ attraction
strength. Below we study the sensitivity to the choice of this parameter.

Using Eqs.~(\ref{scpf})~and~(\ref{dpit1}) one gets the relations
\begin{eqnarray}
&&\Delta f=-(a_N\hsp n_N^2+2a_{N\alpha}n_N\hsp n_\alpha+a_\alpha\hsp n_\alpha^2)
+\frac{b_N}{\gamma+1}\hsp (n_N+\xi\hsp n_\alpha)^{\gamma+2},\label{dfsi}\\
&&\mu_N=\widetilde{\mu}_N-2\hsp (a_N\hsp n_N+a_{N\alpha}\hsp n_\alpha)
+\frac{\gamma+2}{\gamma+1}\hsp b_N\hsp (n_N+\xi\hsp n_\alpha)^{\gamma+1},\label{cpn1}\\
&&\mu_\alpha=\widetilde{\mu}_\alpha-2\hsp (a_{N\alpha}\hsp n_N+a_\alpha\hsp n_\alpha)
+\frac{\gamma+2}{\gamma+1}\hsp b_N\hsp\xi (n_N+\xi\hsp n_\alpha)^{\gamma+1}.\label{cpal1}
\end{eqnarray}

To study the EoS of interacting $\alpha-N$ matter
we choose a certain value of $a_{N\alpha}$ and substitute (\ref{cpn1}) and (\ref{cpal1}) into the condition of chemical equilibrium~(\ref{cce}).
The resulting equation gives allowable states in the ($T,n_N,n_\alpha$) space. Then
from~Eqs.~(\ref{tpwi}),~(\ref{dpit1}), and (\ref{cpn1}) we~determine pressure at different $\mu=\mu_N$ and~$T$.

Before going to numerical results we would like to note that our approach
becomes questionable at high densities of $\alpha$ particles.
Classical Monte-Carlo calculations in the~hard-sphere approximation show~\cite{Mul08} that the transition to
a `solid' phase occurs in a~pure \mbox{alpha}~system at \mbox{$n_\alpha\gtrsim (0.07-0.1)~\textrm{fm}^{-3}$} (in this estimate we assume~the radius of the~$\alpha$~\mbox{nucleus} \mbox{$r_\alpha=1-1.2~\textrm{fm}$}).
Therefore, our results should be considered with caution at baryon densities
\mbox{$n_B\gtrsim 0.3-0.4~\textrm{fm}^{-3}$}.

\section{Results for interacting $\alpha-N$ matter\label{lab6a}}

\subsection{Zero temperature limit\label{lab6}}

Let us consider first the ground state of the $\alpha-N$ matter at zero temperature.
Note that this is the state with $p=0$ and minimal energy per
baryon $\varepsilon/n_B$. Using formulae of the preceding section one can calculate the pressure~$p$, the baryon chemical potential
\bel{cce2}
\mu=\mu_N=\mu_\alpha/4\hsp,
\ee
and the energy per baryon $\varepsilon/n_B=\mu-p/n_B$ at $T\to 0$ as functions
of $n_B$ for different values of the parameter $a_{N\alpha}$.

Our analysis shows that
the results are qualitatively different if this parameter is smaller or larger than
some threshold value $a_*$ (see below~\re{thrc1}). In the region $a_{N\alpha}<a_*$ the~GS of
the $\alpha-N$ mixture corresponds to a pure nucleon matter ($n_\alpha=0$) with $\mu=\mu_0$
and~\mbox{$n_N=n_0$}. Here~$\mu_0$ and $n_0$ are, respectively, the chemical potential and the saturation density of the~equilibrium nucleon matter, introduced in Sec.~\ref{lab2}. In the same interval of~$a_{\alpha N}$, there exists another local minimum of energy per baryon with~$n_N=0$ which corresponds to a~pure alpha matter. This state is metastable because it has a~smaller binding energy as compared to a~pure nucleon matter. These two minima in the~($n_B,\chi$) plane are separated by an energetic barrier.

\begin{figure}[htb!]
\centering
\includegraphics[trim=1.7cm 2.7cm 1.7cm 12.5cm, clip, width=0.48\textwidth]{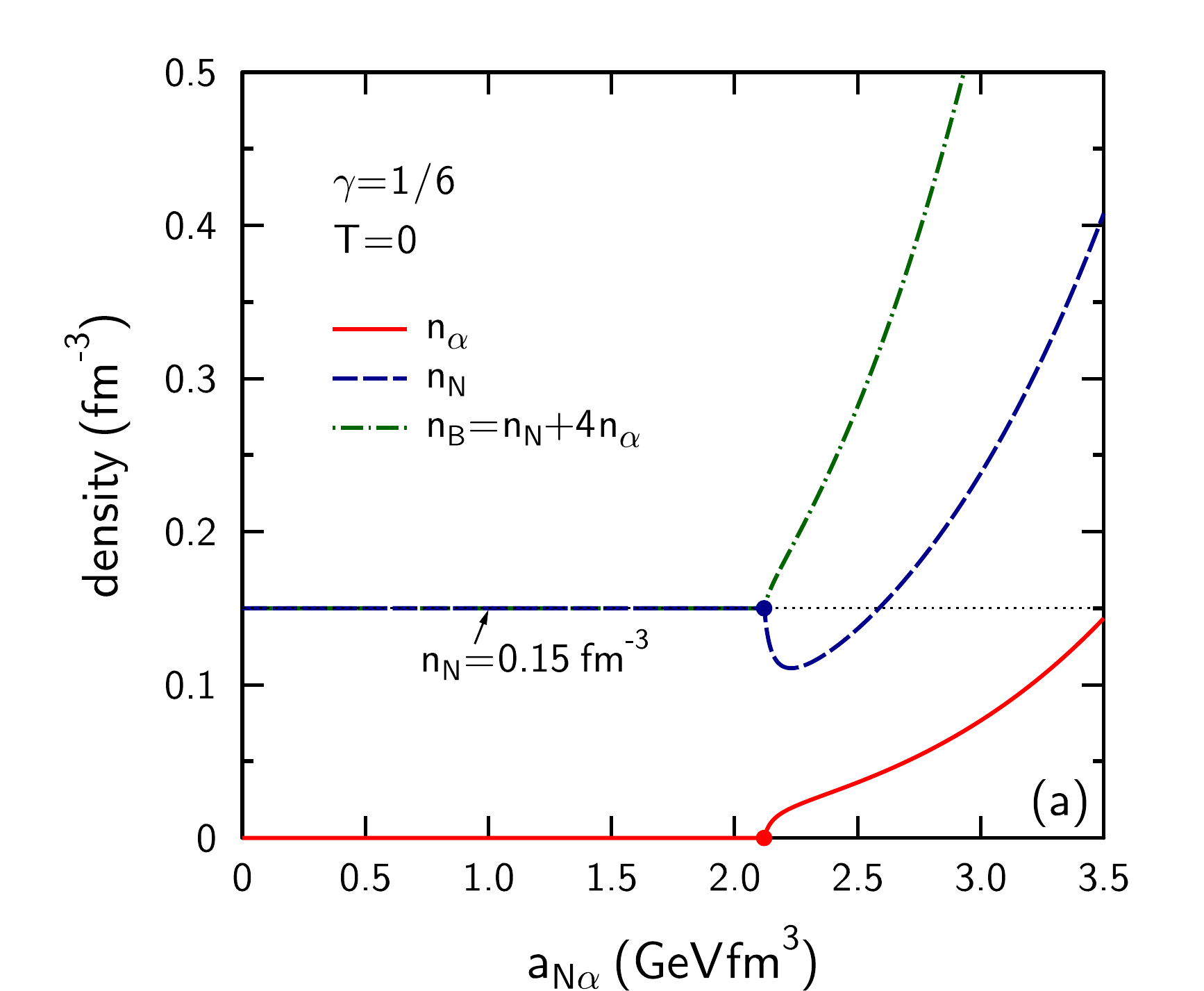}
\includegraphics[trim=1.7cm 2.7cm 1.7cm 12.5cm, clip, width=0.48\textwidth]{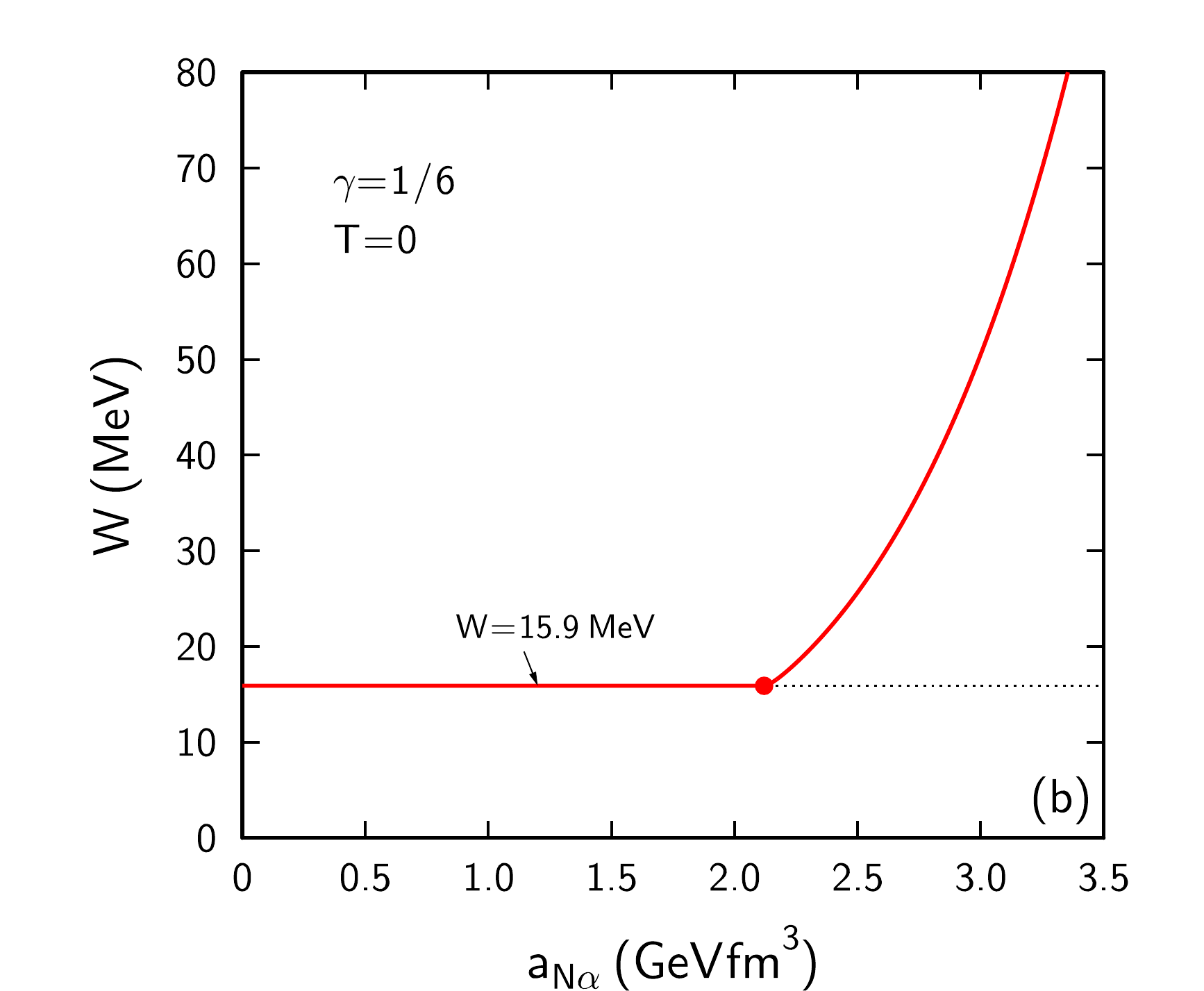}
\caption{
(a) Partial densities as well as baryon density of particles in GS of cold $\alpha-N$ matter
as functions of cross-term coefficient $a_{N\alpha}$. (b) Binding energy per baryon of
cold $\alpha-N$ mixture as the function of~$a_{N\alpha}$. Dots correspond to threshold
value $a_{N\alpha}=a_*$ (see text). All calculations correspond to $\gamma=1/6$.
}\label{fig5}
\end{figure}
Our calculations show that at \mbox{$a_{N\alpha}>a_*$}
the $\alpha-N$ mixture has only one minimum-energy state in the ($n_B,\chi$) plane
and this system becomes stronger bound as compared to a pure nucleon matter.
In~this region the GS is characterized by a nonzero value of $n_\alpha$ and
the~corres\-ponding binding energy $W=m_N-\varepsilon/n_B=m_N-\mu$ increases with $a_{N\alpha}$.

 The threshold value $a_*$ can be found analytically by using formulae of preceding section.
 One should take into account that at zero temperature all $\alpha$'s are Bose-condensed~($\widetilde{\mu}_\alpha=m_\alpha$) and
 the ideal gas pressure $p_\alpha^{\hsp\rm id}=0$. Using these relations
 and formulas of Sec.~\ref{lab4a} one gets the equations
 \begin{eqnarray}
 &&p=p_N^{\hsp\rm id}+\Delta p\hsp(n_N,n_\alpha)=0\hsp ,\label{prgs}\\
 &&\mu_N=E_F(n_N)+U_N(n_N,n_\alpha),\label{cpngs}\\
 &&\mu_\alpha=m_\alpha+U_\alpha(n_N,n_\alpha),\label{cpags}
 \end{eqnarray}
 where $\Delta p,U_N,U_\alpha$ are functions of $n_N,n_\alpha$ defined in
 Eqs.~(\ref{dpit1}),~(\ref{cpn1})--(\ref{cpal1}).

The ground-state values of $\mu,n_B,\chi$ are determined by simultaneously solving
\mbox{Eqs.~(\ref{cce2})--(\ref{cpags})}. They are continuous functions of $a_{N\alpha}$,
so that \mbox{$\mu\to\mu_0$}, \mbox{$n_N\to n_0$}, \mbox{$n_\alpha\to 0$}
 at~\mbox{$a_{N\alpha}\to a_*$}.
 Substituting these limiting values into (\ref{cce2}), (\ref{cpags}) gives
 \bel{thrc}
 m_\alpha+U_\alpha\hsp (n_0,0)=4\hsp\mu_0~~~~~~(a_{N\alpha}=a_*)\hsp.
 \ee
 Solving this equation with respect to $a_*$ gives
 \bel{thrc1}
a_*=\frac{1}{2}\left (\frac{m_\alpha-4\mu_0}{n_0}+
\frac{\gamma+2}{\gamma+1}\,b_N\,\xi\hsp n_0^\gamma\right)\simeq
2.12~\textrm{GeV\hsp fm}^3~~~~(\gamma=1/6)\hsp .
\ee
In the last equality we use numerical values of $b_N,\xi$ obtained in Sec.~\ref{lab2}~and~\ref{lab5}.

Figures~\ref{fig5}\hspm (a) and~(b) show ground-state characteristics of a cold $\alpha-N$ matter
as functions of~$a_{N\alpha}$ for $\gamma=1/6$. One can see that at $a_{N\alpha}>a_*$ the binding
energy and densities $n_\alpha, n_B$ increase monotonically with $a_{N\alpha}$.

Below we present the results for $\gamma=1/6$ and choose the parameter $a_{N\alpha}$ in the interval
\mbox{$a_{N\alpha}<a_*$}, i.e., assume that alphas do not appear in the GS
at~$T\to 0$\hsp. Such an assumption seems to be supported by the nuclear phenomenology.
To~study the sensitivity to $a_{N\alpha}$\hspm, we~compare the results for $a_{N\alpha}=1$ (set~A) and~$1.9$~(set~B)~GeV\hsp fm$^3$.

\subsection{Phase diagram of interacting $\alpha-N$ matter\label{lab7}}

In this section we consider the EoS of the chemically equilibrated $\alpha-N$ mixture
at nonzero temperatures. We apply explicit relations for pressure and free energy
derived in Sec.~\ref{lab5}. By solving~\re{cce} one can find allowable states of matter
in the $(T,\mu,p)$ or~$(T,n_N,n_\alpha)$ space. Stability of such states is~studied by
calculating the sign of the determinant in~\re{stcon}.
\begin{figure}[htb!]
\centering
\includegraphics[trim=1.7cm 1.9cm 1.7cm 1.4cm, clip, width=0.8\textwidth]{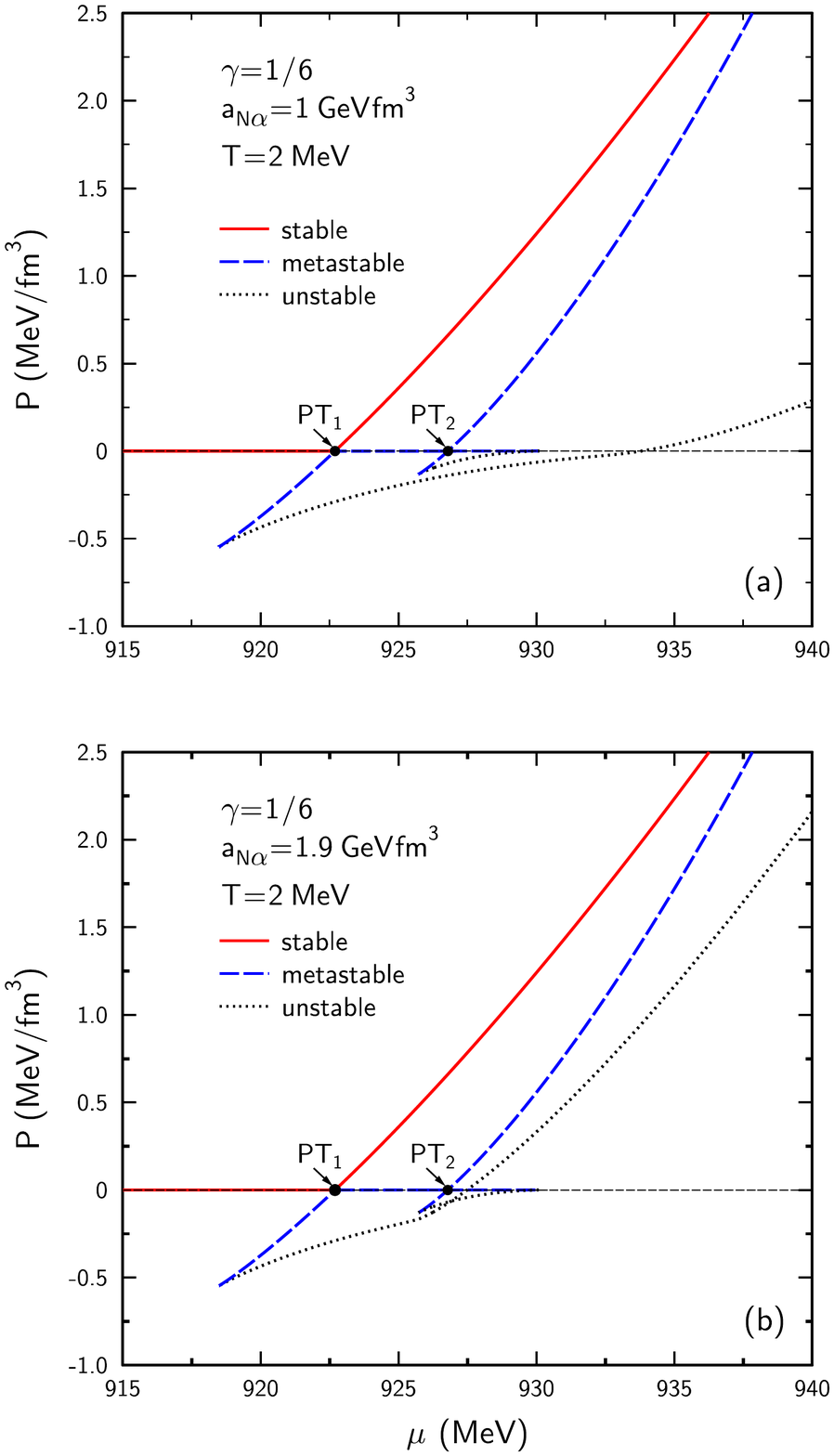}
\caption{
The isotherm $T=2~\textrm{MeV}$ of $\alpha-N$ matter in $(\mu,p)$ plane for
the parameter sets A~(a) and~B~(b). The stable, metastable and unstable
parts of isotherm are shown by the solid, dashed and dotted lines, respectively.
The dots PT$_1$ and PT$_2$ show positions of stable and meta\-stable~LGPT, respectively.
}\label{fig6}
\end{figure}

\begin{figure}[htb!]
\centering
\includegraphics[trim=1.7cm 1.5cm 1.7cm 1.8cm, clip, width=0.8\textwidth]{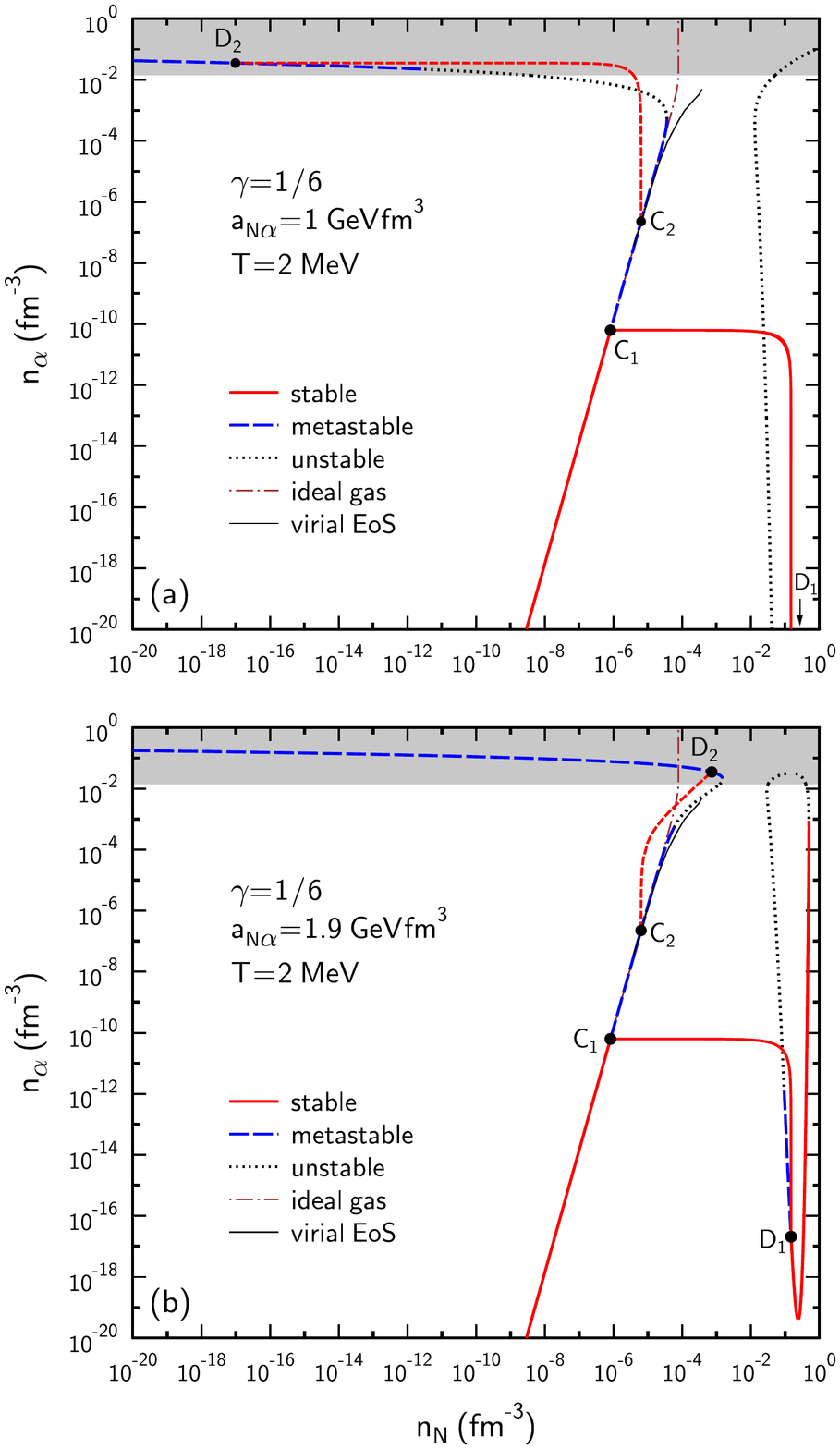}
\caption{
The isotherm $T=2~\textrm{MeV}$ in $(n_N,n_\alpha)$ plane for the same parameters as
in Fig.~\ref{fig6}. The~BEC region is shown by shading.
The dash--dotted line corresponds to ideal $\alpha-N$ gas.
Lines $C_1D_1$ and~$C_2D_2$ correspond to mixed--phase states of~PT$_1$ and~PT$_2$,
respectively. The thin solid line represents the isotherm $T=2~\textrm{MeV}$ from Ref.~\cite{Hor06}.
Note that binodal point $D_1$ in upper panel lies below the plotted region.
}\label{fig7}
\end{figure}

Figure~\ref{fig6} represents the isotherm $T=2~\textrm{MeV}$ in the plane $(\mu,p)$\hsp .
Lower and upper panels correspond to sets A and B, respectively.
Unstable parts of the isotherm are shown by dotted lines. It is interesting that only these parts exhibit significant changes in the transition between sets A and B.
According to the Gibbs rule, intersection points of (meta)stable branches of pressure as functions of $\mu$ correspond to phase transitions (PT).
As one can see from Fig.~\ref{fig6}, there are two PT at $T=2~\textrm{MeV}$. Their characteristics,
in particular, critical values of the baryon chemical potential $\mu_c$ are given in Table~\ref{tab3}.

\begin{figure}[htb!]
\centering
\includegraphics[trim=1.7cm 1.5cm 1.7cm 1.8cm, clip, width=0.8\textwidth]{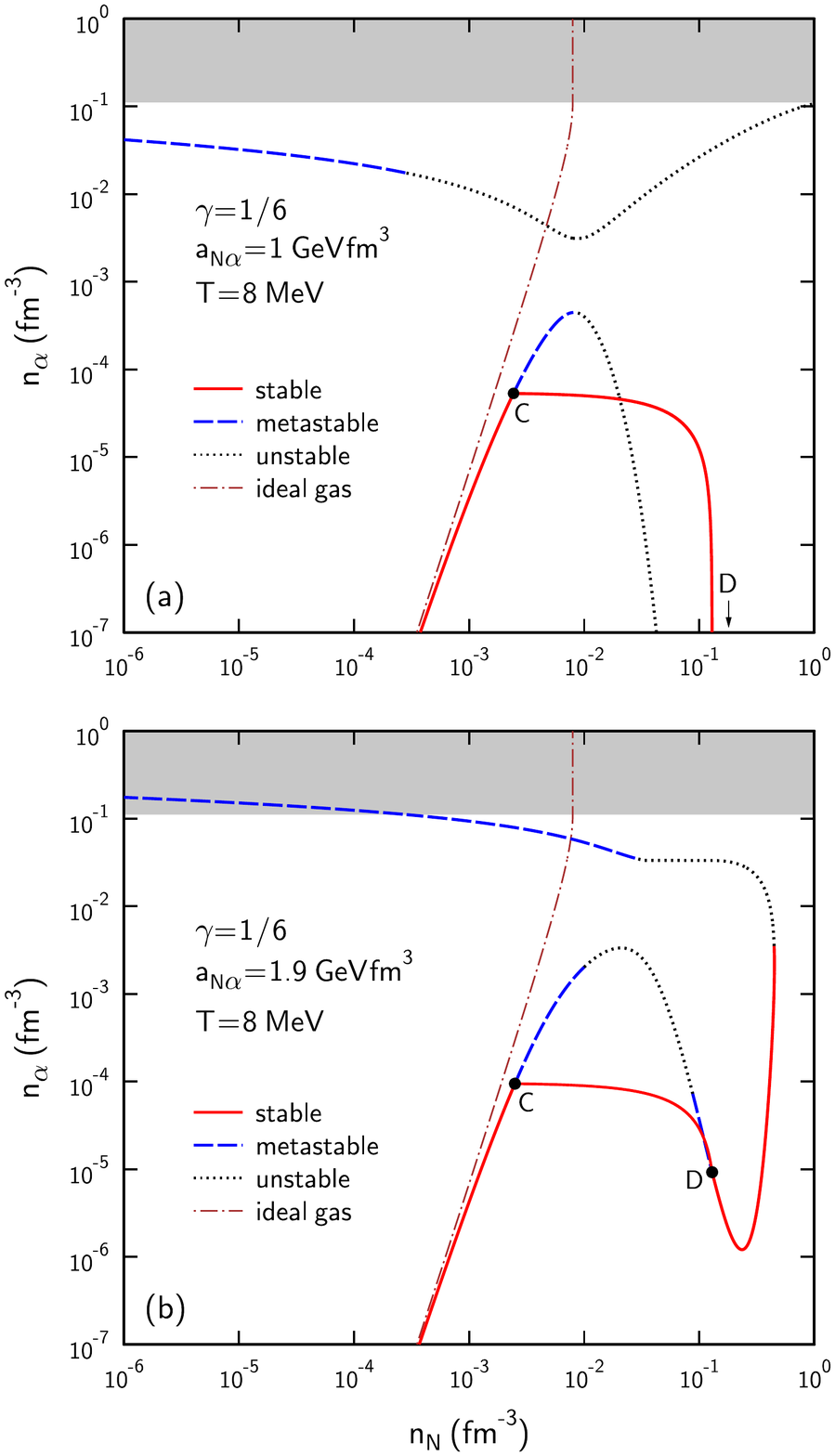}
\caption{
Same as Fig.~\ref{fig7} but for $T=8~\textrm{MeV}$. Note that only one (stable) LGPT
exists at this temperature.
}\label{fig8}
\end{figure}

The first transition,~PT$_1$, occurs at a smaller baryon chemical potential as compared
to~PT$_2$\hsp\footnote
{
Note that the slope of pressure as a function of $\mu$ equals the baryon density $n_B$.
Therefore, jumps of~the~pressure slopes at points~PT$_1$ and PT$_2$
in Fig.~\ref{fig6}, correspond to nonzero jumps of $n_B$\hsp .
}.
As~a~consequence, states on the dashed lines have smaller pressure (i.e. larger thermodynamic potential $\Omega=-p\hsp V$) as compared to states with the same $\mu$ on the solid curve. It is well-known, that states with smaller pressure are thermodynamically less favorable~\cite{Lan75,Gre97}.
This shows that states on dashed lines (including mixed-phase states of PT$_2$)
are metastable. In Figs.~\ref{fig6}--\ref{fig8} stable (favorable) states are represented by the solid lines and metastable (unfavorable) states are indicated by the dashed lines. The unstable states with maximum values of~$\Omega$ are shown by dotted lines.

Figures~\ref{fig7}\hspm (a) and~(b) represent the same isotherm $T=2~\textrm{MeV}$, but
in the $(n_N,n_\alpha)$ plane.
A~strong sensitivity to $a_{N\alpha}$ is clearly visible in this representation.
By shading we~show the region of~BEC
$n_\alpha>n_{\alpha}^*\simeq 0.014~\textrm{fm}^{-3}$ (see Sec.~\ref{lab1}).
For both sets of parameters we do not find any stable states of the $\alpha-N$ matter with large fraction of alphas at den\-sities~$n_N\gtrsim 10^{-2}~\textrm{fm}^{-3}$. One may say that the model imitates the Mott effect~\cite{Roe82}, i.e., predicts a suppression of nuclear clusters at large baryon densities.
On the other hand, the model predicts metastable states, where $\alpha$ particles are more abundant than nucleons (see,~e.g., upper parts of Figs.~\ref{fig7},~\ref{fig8} where the dashed lines enter the shaded area).

\hspace*{-1cm}\begin{table}[htb!]
\caption{Characteristics of LGPT for $\alpha-N$ matter
at $T=2~\textrm{MeV}$ (set~B)\hsp .}
\vspace*{2mm}
\label{tab3}
\footnotesize
\begin{tabular}{|c|c|c|c|c|c|c|c|c|c|}
\hline
&&\multicolumn{4}{c|}{binodal point $C$}&\multicolumn{4}{c|}{binodal point $D$}\\
\cline{2-10}
&~$\mu_c-m_N$~&~$n_N$ &~$n_{\alpha}$ &~$n_B$
&~$\chi$ &~$n_N$  &~$n_{\alpha}$
&~$n_B$ &~$\chi$\\[-2mm]
&~$\left(\textrm{MeV}\right)$~&~$\left(\textrm{fm}^{-3}\right)$ &~$\left(\textrm{fm}^{-3}\right)$ &~$\left(\textrm{fm}^{-3}\right)$
&~&~$\left(\textrm{fm}^{-3}\right)$ &~$\left(\textrm{fm}^{-3}\right)$
&~$\left(\textrm{fm}^{-3}\right)$ &~\\
\hline
~PT$_1$~&~$-16.2$~&~$8.2\cdot 10^{-7}$ &~$6.3\cdot 10^{-11}$ &~$8.2\cdot 10^{-7}$ &~$3.1\cdot 10^{-4}$
&~$0.15$ &~$2.1\cdot 10^{-17}$ &~$0.15$ &~$1.4\cdot 10^{-16}$\\
\hline
PT$_2$ &~$-12.1$~&~$6.4\cdot 10^{-6}$ &~$2.3\cdot 10^{-7}$ &~$7.3\cdot 10^{-6}$ & 0.12 &~$7.3\cdot 10^{-4}$
&~$3.5\cdot 10^{-2}$ &~0.14 & 1.0\\
\hline
\end{tabular}
\end{table}

Points $C_{\hsp i}$ and $D_{\hsp i}$ in Fig.~\ref{fig7} and Table~\ref{tab3} are the binodal points (i.e. boundaries of the liquid-gas MP) for the phase transition PT$_i~(i=1,2)$. Coordinates of such points in
the~$(n_N,n_\alpha)$ plane are determined from the Gibbs conditions of phase equilibrium:
\begin{eqnarray}
&&p\left(T,n_N^{\mbox{\footnotesize $(C)$}},n_\alpha^{\mbox{\footnotesize $(C)$}}\right)=
p\hsp \left(T,n_N^{\mbox{\footnotesize $(D)$}},n_\alpha^{\mbox{\footnotesize $(D)$}}\right)=
p_{\hsp c}\hsp,\label{pgcpe}\\
&&\mu_N\left(T,n_N^{\mbox{\footnotesize $(C)$}},n_\alpha^{\mbox{\footnotesize $(C)$}}\right)=
\mu_N\left(T,n_N^{\mbox{\footnotesize $(D)$}},n_\alpha^{\vspace*{-1mm}\mbox{\footnotesize $(D)$}}\right)
=\mu_{\hsp c}\hsp,\label{mgcpe}
\end{eqnarray}
where we omit indices $i$\hsp . Characteristics of binodal points for the parameter set~B are given in Table~\ref{tab3}\hsp\footnote
{
Calculation with set~A gives a very small value (about
$10^{-74}~\textrm{fm}^{-3}$) for the alpha-particle density at the~binodal point $D_1$\hsp.
The latter lies far below the horizontal axis in Fig.~\ref{fig7}\hspm (a).
}.
We have checked that at $T=2~\textrm{MeV}$ the nucleon densities at points $C_1$ and $D_1$ are close
to binodal densities of a pure nucleon matter (see Sec.~\ref{lab2}). The same conclusion
is valid for the alpha-particle densities at points $C_2$ and $D_2$: they are close to the binodal densities
obtained for a pure alpha matter in Sec.~\ref{lab3}. Note that for both parameter sets
point $D_2$ lies in the~BEC region.

The solid and short-dashed lines $C_1D_1$ and $C_2D_2$ in Fig.~\ref{fig7} correspond to the
MP states for~PT$_1$ and PT$_2$\hsp , respectively. Coordinates of these states
in the ($n_N,n_\alpha$) plane and the~\mbox{volume} fraction of the liquid phase, $\lambda$, satisfy
the relations (as above, we omit the~phase transition index)
\bel{vflp}
\lambda=\frac{n_N-n_N^{(C)}}{n_N^{(D)}-n_N^{(C)}}=
\frac{n_\alpha-n_\alpha^{(C)}}{n_\alpha^{(D)}-n_\alpha^{(C)}}\,.
\ee
One can see that the MP states lie on the straight lines in the ($n_N,n_\alpha$) plane. However,
one can hardly recognize this linear dependence in~Fig.~\ref{fig7} because of the double-logarithmic scale used in this plot.

As one can see from Table~\ref{tab3} and Fig.~\ref{fig7}, the mass fraction of alphas,~$\chi$, is relatively small for the MP states of PT$_1$\hsp, but it is rather large for the
transition PT$_2$\hsp. As has been already mentioned, the phase transition PT$_2$ is in fact
metastable. Nevertheless, we believe that it can be observed in dynamical processes (like heavy-ion collisions) by selecting states
with large relative abundances of $\alpha$'s. The same statement can be made for
BEC states (see the dashed lines in the shaded regions). Indications of enhanced production of so-called $\alpha$-conjugate nuclei have been observed recently in intermediate-energy nuclear collisions~\cite{Sch17}.

The results obtained within a virial approach~\cite{Hor06} are shown in~Fig.~\ref{fig7}
by thin solid lines. This approximation can be considered as reasonable only at low densities. Note that the~quantum--statistical and phase transition effects are disregarded in such a model.
Never\-theless, from Fig.~\ref{fig7} one can conclude that calculations with set~B are in better agreement with the results of Ref.~\cite{Hor06}. Presumably, this parameter set is preferable as compared to set~A.

Figures~\ref{fig8}\hspm (a) and~(b) show the isotherm $T=8~\textrm{MeV}$
in the $(n_N,n_\alpha)$ plane, again for the~parameter sets A and B. At this temperature
only one, stable LGPT remains. One can see a~significant change in shape of the
isotherm as compared to the case $T=2~\textrm{MeV}$ considered in Fig.~\ref{fig7}.
\begin{table}[htb!]
\caption{Characteristics of phase transitions in $\alpha-N$ matter for parameter sets A and B.}
\vspace*{2mm}
\label{tab4}
\footnotesize
\begin{tabular}{|c|c|c|c|c|c|c|c|c|c|}
\hline
&\multicolumn{4}{c|}{PT$_1$~~}&\multicolumn{5}{c|}{PT$_2$}\\
\cline{2-10}
&~$T_{\rm CP}$~&~$\mu_{\hspm\rm CP}-m_N$~&~$n_{B\rm CP}$~&~$\chi_{\rm CP}$~&
~$T_K$~&~$\mu_K-m_N$ &~$n_{BK}$ &~$\chi_K$ &$T_{\rm TP}$\\[-2mm]
&~$\left(\textrm{MeV}\right)$~&~$\left(\textrm{MeV}\right)$~&~$\left(\textrm{fm}^{-3}\right)$~&~
&~$\left(\textrm{MeV}\right)$~&~$\left(\textrm{MeV}\right)$~&~$\left(\textrm{fm}^{-3}\right)$
&~&$\left(\textrm{MeV}\right)$\\
\hline
~set~A~&~$15.4$~&~$-31.7$~&~$4.8\cdot 10^{-2}$ &~$2.5\cdot 10^{-4}$
&~$7.6$ &~$-14.3$ &~$(1.2-2.6)\cdot 10^{-2}$ &~$0.14-1.0$ &~$3.54$
\\
\hline
~set~B~&~$14.7$~&~$-30.3$~&~$5.3\cdot 10^{-2}$ &~$6.9\cdot 10^{-2}$
&~$4.6$ & $-13.2$ &~$1.3\cdot 10^{-3}-10^{-1}$&~$0.46-0.86$ &~$3.37$
\\
\hline
\end{tabular}
\end{table}

\begin{figure}[htb!]
\centering
\includegraphics[trim=1.7cm 7cm 1.7cm 9cm, clip, width=0.48\textwidth]{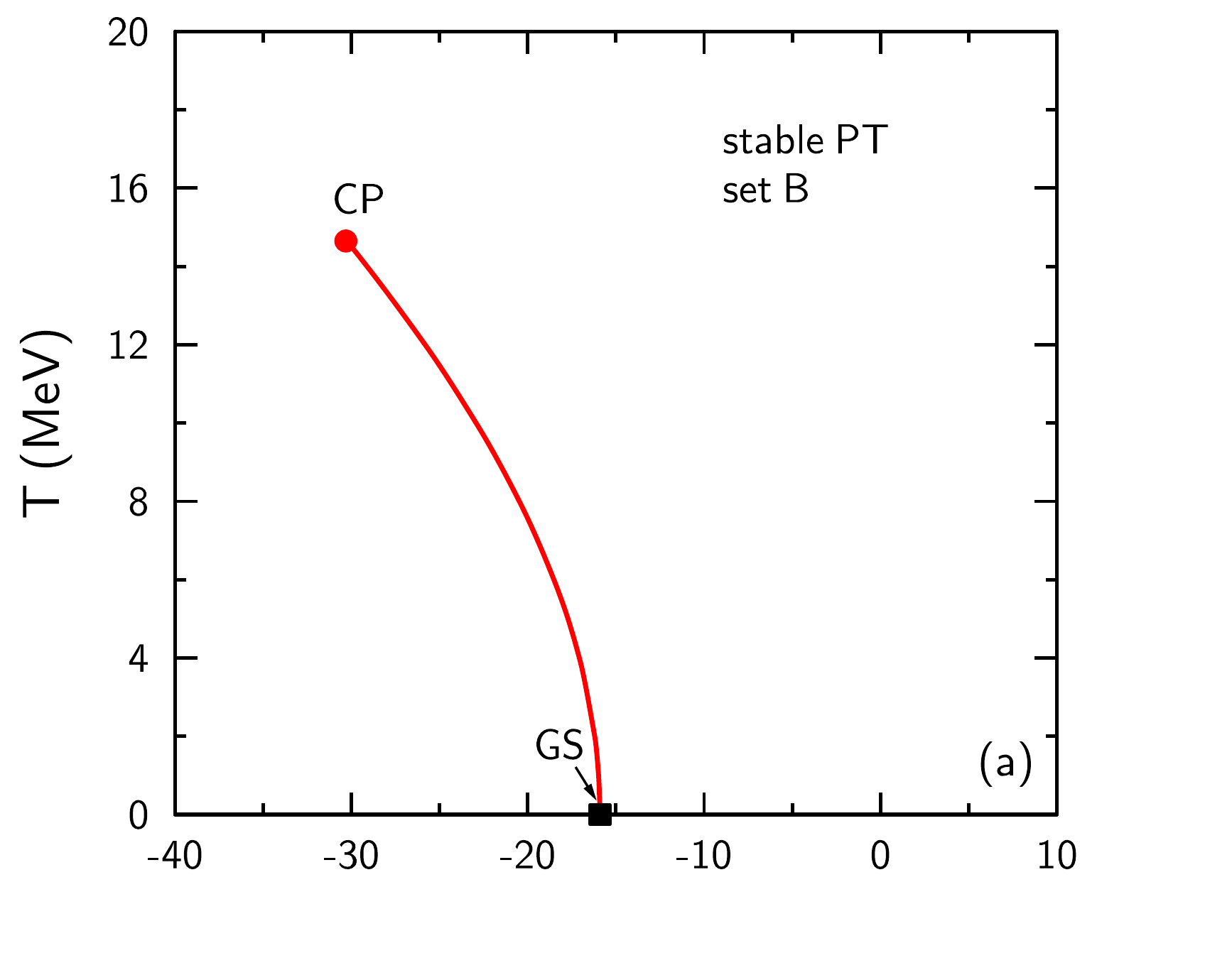}
\includegraphics[trim=1.7cm 6.8cm 1.7cm 9cm, clip, width=0.48\textwidth]{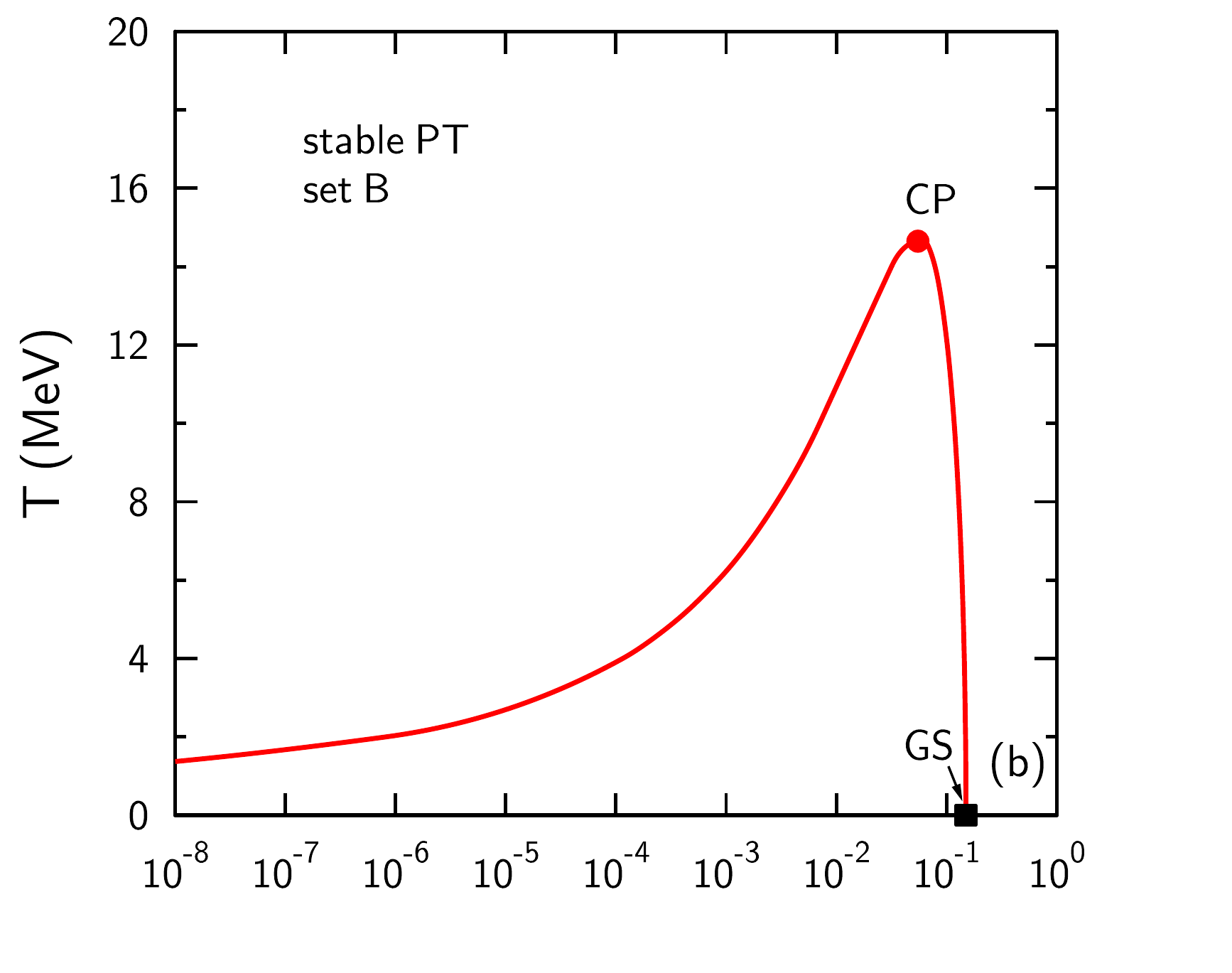}
\includegraphics[trim=1.7cm 7cm 1.7cm 9cm, clip, width=0.48\textwidth]{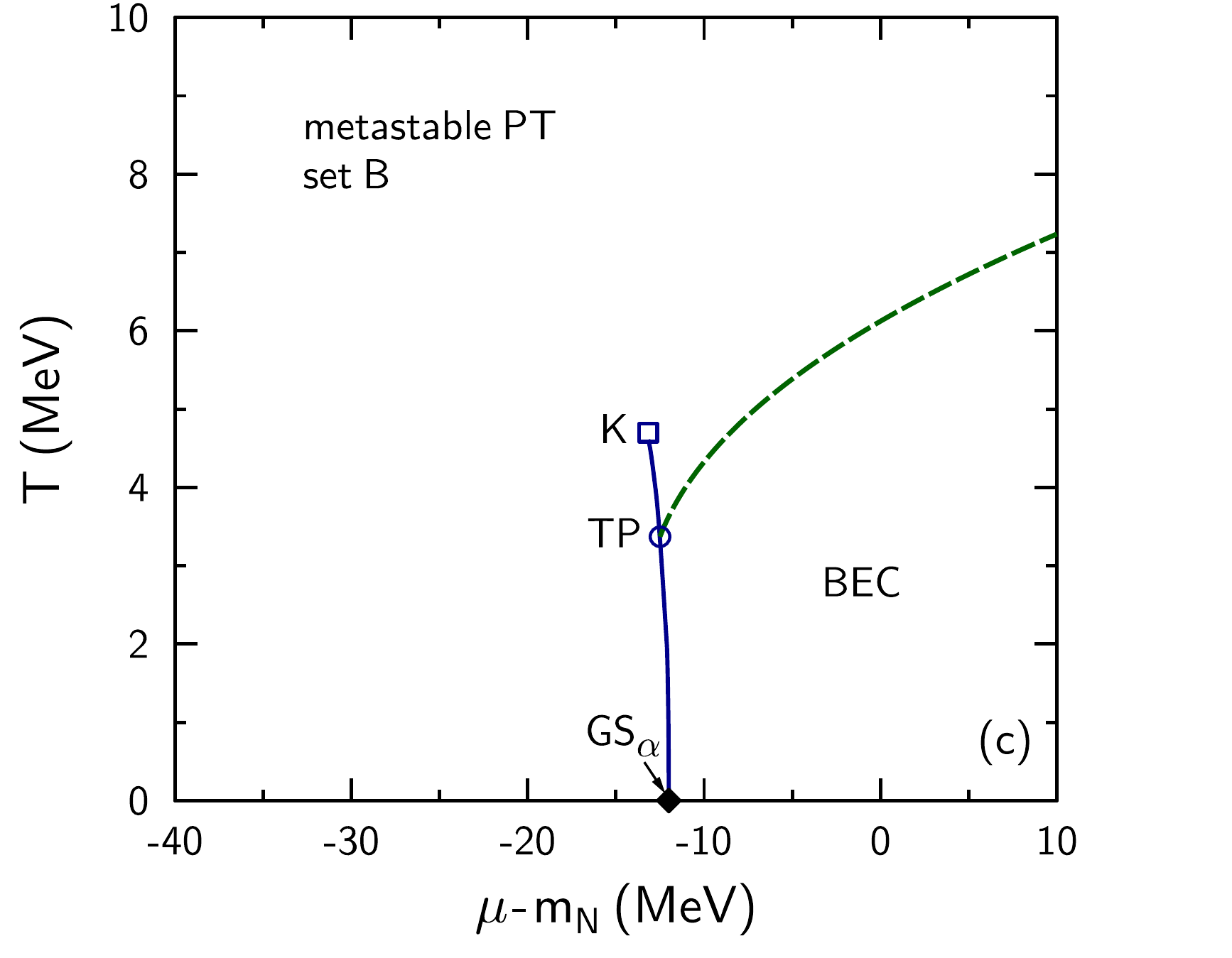}
\includegraphics[trim=1.7cm 6.8cm 1.7cm 9cm, clip, width=0.48\textwidth]{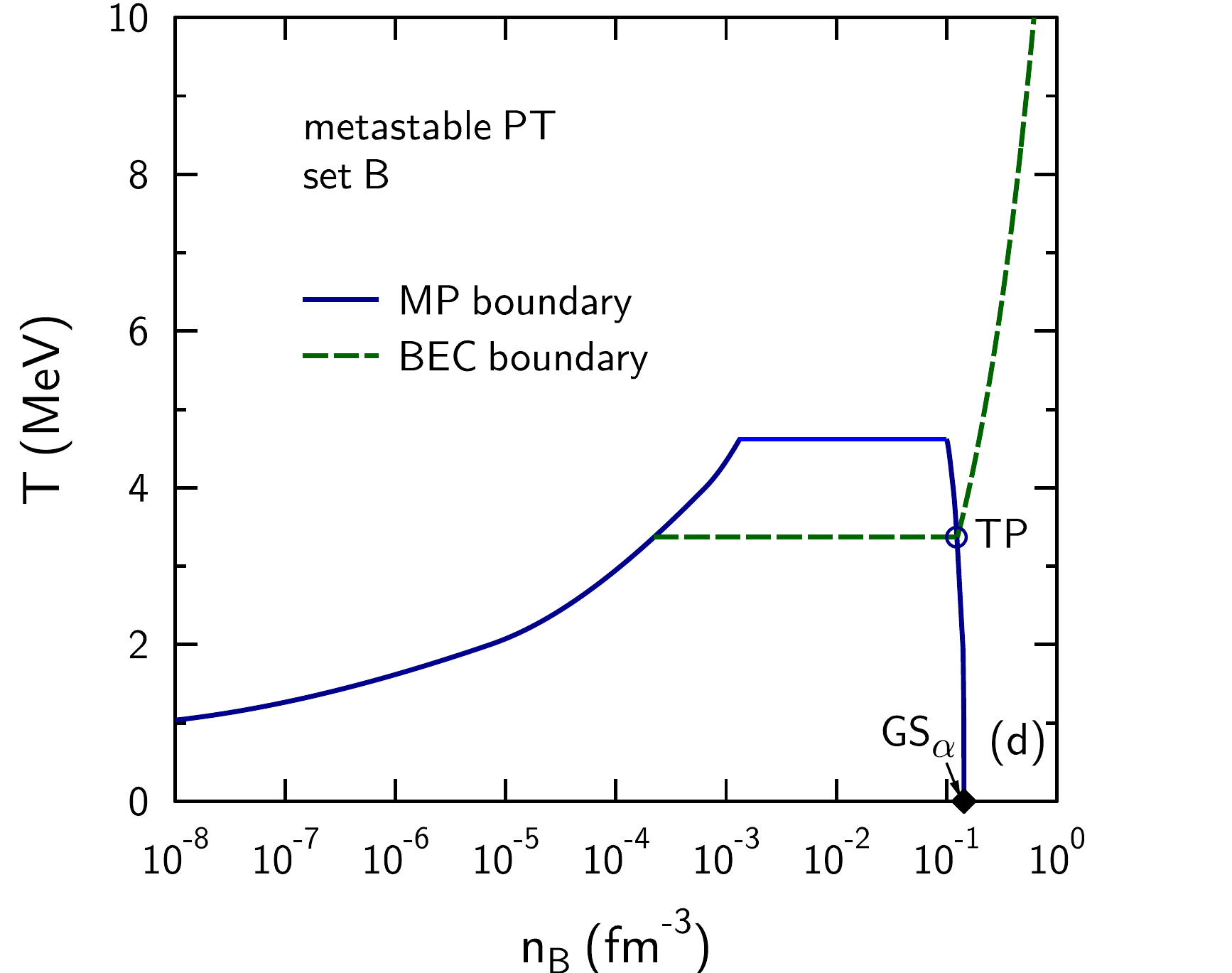}
\caption{
Left panels: critical lines of stable (a) and metastable (c) PT of $\alpha-N$ matter in $(\mu,T)$ plane.
Right panels: boundaries of MP for stable (b) and metastable (d) PT of $\alpha-N$ mixture
in~$(n_B,T)$ plane. All calculations correspond to parameter set~B.
Full circles in (a) and (b) show positions of critical point. The~dashed
lines in (c) and (d) represent boundaries of BEC region. The~open square (circle)
marks the end (triple) point of the metastable~PT. Full squares and diamonds show,
respectively, the GS positions for pure nucleon and pure alpha matter.
}\label{fig9}
\end{figure}
Further increase of $T$ leads to disappearance of the LGPT.
This takes place at~$T>T_{\rm CP}$ where $T_{\rm CP}$ is the temperature
of the critical point. Similar to pure nucleon and alpha matter, we determine
characteristics of this point by simultaneously solving the
equa\-tions~$(\partial\hsp p/\partial\hsp n_B)_T=0$ and
$(\partial^{\hsp 2} p/\partial^{\hsp 2}n_B)_T=0$\,. Our analysis shows that the metastable transition PT$_2$ disappears 'abruptly'
at some temperature, $T_K$ which is smaller than $T_{\rm CP}$. Note that there is still
a~nonzero baryon density jump at $T=T_K$ (see Fig.~\ref{fig9}\hspm (d))\hsp\footnote
{
This means that at $T>T_K$ there are no additional intersections between the pressure branches in
the~($\mu,p$) plane except the point PT$_1$. We found, that pressure slopes on both sides of the point PT$_2$ still differ when~$T\to T_K$. On the other hand, the density jump disappears for the transition PT$_1$ when~\mbox{$T\to T_{\rm CP}$}.
}.

Table~\ref{tab4} gives characteristics of the critical point of the PT$_1$ as well as those for
the end point~$K$ of the PT$_2$\hspm . One can see that position of the critical point CP is not very sensitive to the parameter~$a_{N\alpha}$. On the other hand, characteristics of PT$_2$ are stronger modified in the transition between sets~A and~B.

A more detailed information is given in Figs.~\ref{fig9}\hspm (a)--(d)  which represent
the phase diagrams of the $\alpha-N$ matter in the $(\mu,T)$ and $(n_B,T)$ planes. Qualitatively, the critical line of the~metastable PT in the ($\mu,T$) plane is similar to that for a pure $\alpha$ matter (see Fig.~\ref{fig4}\hspm (a)). Note however, that the end point $K$ can not be regarded as a critical point.
The full square and diamonds in Figs.~\ref{fig9} mark, respectively, the ground states of one-component systems composed of nucleons or alphas. These states coincide with boundaries
of~critical lines on the~axis $T=0$.

\begin{figure}[htb!]
\centering
\includegraphics[width=0.75\textwidth]{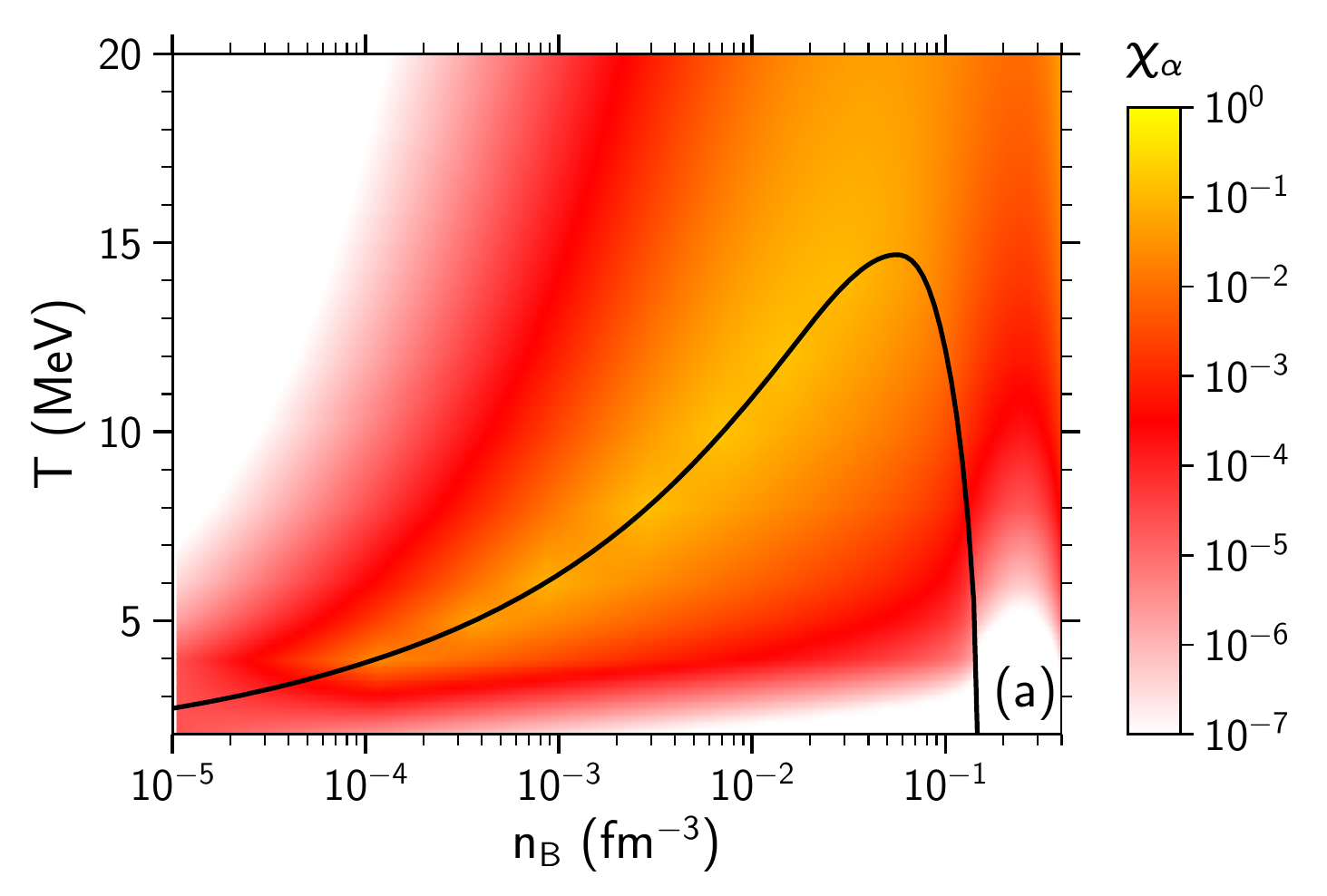}
\includegraphics[width=0.75\textwidth]{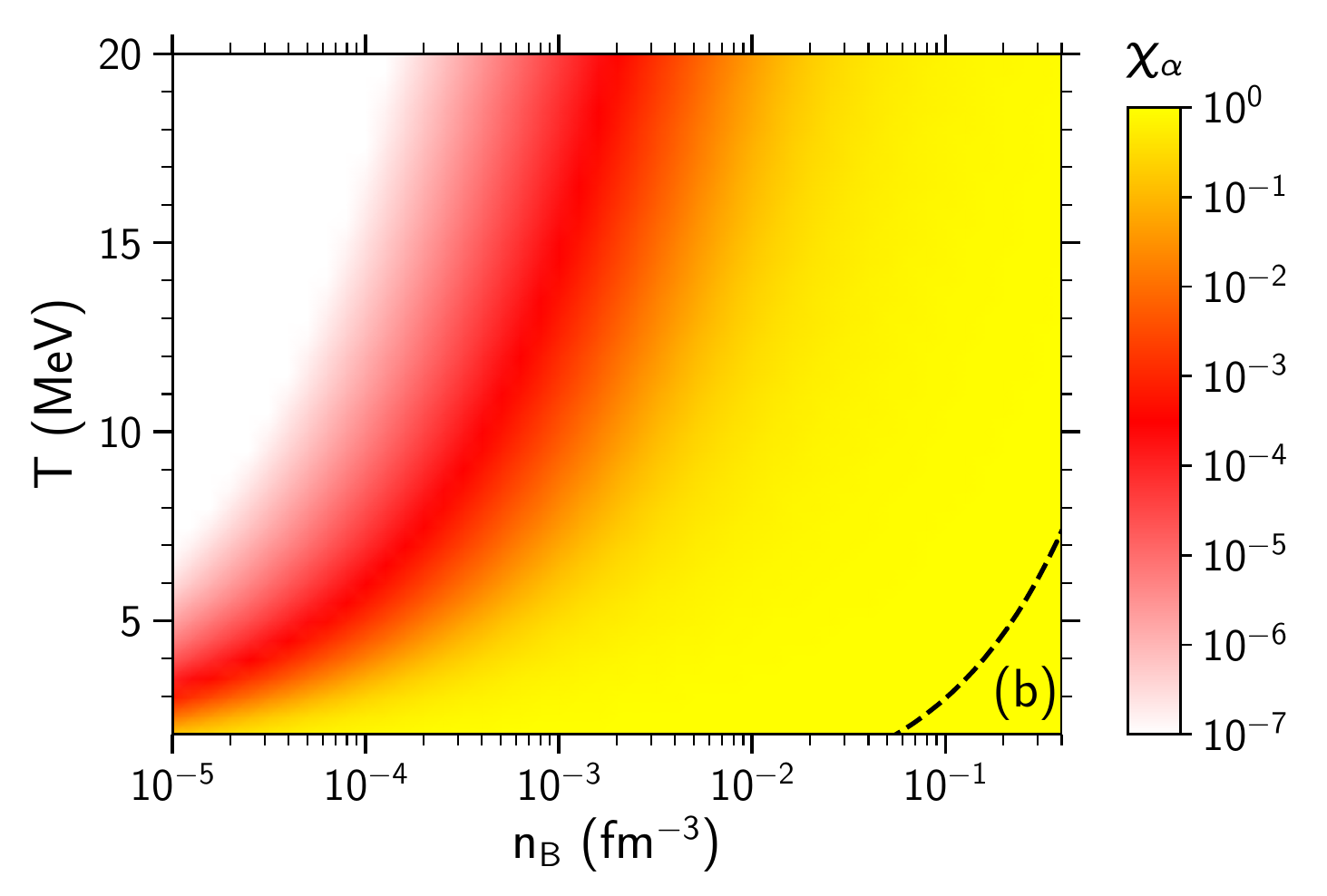}
\caption{
(a) Contour plot of the mass fraction of $\alpha$'s in $\alpha-N$ matter for parameter set B.
The~MP boundary is shown by the solid line. (b) Same as in upper panel, but for ideal $\alpha-N$ gas.
The~dashed line represents the BEC boundary.
}\label{fig10}
\end{figure}
The contour plot of the mass fraction $\chi$ in the $(n_B,T)$ plane is shown in
Fig.~\ref{fig10}\hspm (a) for the~parameter set B.
In~this calculation we take into account only stable states of the $\alpha-N$ matter.
One can see that maximum values $\chi\sim 0.1-0.2$ are reached near the~left boundary of LGPT\hsp\footnote
{
Note that much larger relative abundances of $\alpha$'s and even their BEC can be reached by selecting metastable states of the $\alpha-N$ matter.
}.
At~fixed temperature $\chi$ decreases with $n_B$ in the MP \mbox{region}.
It is interesting that similar nonmonotonic density behavior of $\chi$ was also
predicted in Refs.~\mbox{\cite{Vov17a,Lat91,Hem10,Pai18,Lal18}}. We would like to emphasize that the model gives qualitatively
\mbox{different} results as compared to the ideal $\alpha-N$ gas where the mass fraction of $\alpha$'s increases monotonically with~$n_B$\hsp (see~Fig.~\ref{fig10}\hspm (b)).

\section{Conclusions and outlook\label{lab8}}

In this paper we have analyzed the EoS and phase diagram of the chemically
equilibra\-ted~$\alpha-N$ matter. Our approach simultaneously takes into
account the quantum-statistical effects as well as liquid-gas phase transitions.
We apply Skyrme-like parametrizations of inte\-raction terms as functions of particle
densities. The model parameters were chosen by using the ground-state characteristics of
a pure nucleon and pure alpha matter at zero temperature. We investigate stability
of the $\alpha-N$ mixture with respect to density fluctuations. The regions
of possible phase transitions have been studied for different choices of model
parameters. At low enough temperatures two LGPT are found, where one is stable and other
is metastable. It is demonstrated that the phase transition effects are important
even for dilute states of the $\alpha-N$ matter. A strong suppression of alpha-cluster
abundance is found at large nucleon densities. On the other hand, nucleon fractions
are relatively small for metastable states with Bose--Einstein condensation of alphas.

The results of this paper may be used for studying nuclear cluster production
in heavy-ion reaction as well as in astrophysics.
To analyze dynamical processes in nuclear collisions, it~would be interesting
to calculate not only isotherms but also trajectories of constant entropy per baryon.
Then one can study a possibility to reach the metastable states of $\alpha$~condensation
in the course of isentropic expansion of excited matter produced in a heavy-ion collision.

In the present paper we use parametrizations of mean-fields which predict two separated minima
of the energy per baryon of cold $\alpha-N$ matter. These minima correspond to the~ground states of
pure nucleon and pure $\alpha$ matter. Another possibility, where the nuclear matter has only one
ground state composed of nucleons with a small admixture of $\alpha$'s, will be considered
in the subsequent paper.

In the future, we are going to apply our approach for studies of clusterized isospin-asymmetric matter as expected in compact stars and their merges. More realistic calculations can be made by taking into account the Coulomb interactions as well as contributions of other light and heavy clusters. The results of this paper may be useful for investigating not only equilibrium, but also nonequilibrium mixtures of nucleons and nuclear clusters. We~think that the present model can be also used to study properties of binary mixtures of fermionic atoms and bosonic molecules, like $H+H_2$ or $D+D_2$\hsp.

\begin{acknowledgments}
I.N.M. acknowledges the financial support of the Helmholtz International Center for FAIR \mbox{research}. The work of M.I.G. is supported by the Alexander von Humboldt Foundation and by
the Goal-oriented Program of National the Academy of Sciences of~Ukraine.
L.M.S.~thanks for the support from the Frankfurt Institute for Advanced Studies.
H.St.~appre\-ciates the support from J.~M.~Eisenberg Laureatus chair.
\end{acknowledgments}

\appendix
\section{Thermodynamic functions of ideal nucleon and~alpha gas}
\label{app-A}

Let us consider the case $\mu_\alpha<m_\alpha$
when the density of Bose-condensed alphas $n_{bc}=0$\hsp.
In the lowest order in $T/m_i~(i=N,\alpha)$ one gets from Eqs.~(\ref{prfb}), (\ref{denfb}) the
relations~\cite{Pol17}
\bel{A1}
n_i\simeq\frac{g_i}{\lambda_i^3(T)}\,\Phi^{\pm}_{3/2}\hspace*{-3pt}
\left(\frac{\mbox{$\mu_i-m_i$}}{T}\right),
~~p_i^{\,\rm id}\simeq\frac{g_iT}{\lambda_i^3(T)}\,\Phi^{\pm}_{5/2}\hspace*{-3pt}
\left(\frac{\mu_i -m_i}{T}\right)~~~~(T\ll m_i)\hsp.
\ee
Here upper and lower signs correspond to \mbox{$i=N$} and \mbox{$i=\alpha$}, respectively,
$\lambda_{\hsp i}\hsp (T)$ is the thermal wave length introduced in Sec.~\ref{lab1}, and
\bel{A2}
\Phi^{\pm}_{\beta}(\eta)\equiv\frac{1}{\Gamma(\beta)}\int\limits_0^\infty\frac{x^{\beta-1}dx}
{\ds e^{x-\eta}\pm 1}\,,
\ee
where $\Gamma(\beta)$ is the gamma function\hsp\footnote
{
Note that $\Phi^{\pm}_{\beta}(\eta)=\mp Li_\beta(\mp e^\eta)$ where
$Li_\beta(x)=\sum_{k=1}^{\infty}x^k\hsp k^{-\beta}$ is the polylogarithm of the $\beta$-th order.
}.
For $\eta\leqslant 0$ one can use the decomposition in powers of fugacity:
$\Phi^{\pm}_{\beta}(\eta)=\sum_{k=1}^{\infty}(\mp 1)^{(k+1)}k^{-\beta}\exp{(\eta\hsp k)}$\hsp.
At $\eta=0$ functions (\ref{A2}) are expressed through the Riemann zeta function $\zeta(\beta)$:
\bel{A3}
\Phi^{-}_{\beta}(0)=\zeta(\beta),~~\Phi^{+}_{\beta}(0)=
\left(1-2^{\hsp 1-\beta}\right)\zeta(\beta)\hsp .
\ee

The classical Boltzmann approximation corresponds to the limit $\mu_i-m_i\to-\infty$.
Using the approximate relation $\Phi^{\pm}_{\beta}(\eta)\simeq e^{\hsp\eta}$ at $\eta\to -\infty$
one gets, instead of \re{A1}, much simpler relations
\bel{A4}
n_i\simeq \frac{g_i}{\lambda_i^3(T)}\exp{\left(\frac{\mu_i-m_i}{T}\right)},
~~~p_i^{\,\rm id}\simeq n_iT~~~(\mbox{$n_i\lambda^3_i\ll g_i$}).
\ee

\end{document}